\documentclass[aps,%
reprint,
superscriptaddress,
amsmath,amssymb,floatfix,
prl
]{revtex4-2}
\usepackage{lipsum}
\usepackage{hhline}
\usepackage{upgreek}
\usepackage{orcidlink}

\usepackage{graphicx}
\usepackage{comment}
\usepackage{bm}
\usepackage{xcolor}
\usepackage{makecell}
\usepackage{amsmath,amsfonts,amssymb}
\DeclareMathAlphabet{\mathbbold}{U}{bbold}{m}{n}
\definecolor{darkblue}{rgb}{0,0,0.6}
\usepackage{hyperref}
\hypersetup{colorlinks,linkcolor=darkblue,citecolor=darkblue,urlcolor=darkblue}

\usepackage[normalem]{ulem}

\begin{document}

\title{Odd Darcy Law: Odd Viscosity and Edge Currents in Porous Media}

\author{Raphaël Maire,\orcidlink{0009-0007-7530-9626}}
\email{maire@ub.edu}
\affiliation{Departament de Física de la Mat\`eria Condensada, Universitat de Barcelona, Martí Franqu\`es 1, 08028 Barcelona, Spain}
\author{Ignacio Pagonabarraga,\orcidlink{0000-0002-6187-5025}}
\email{ipagonabarraga@ub.edu}
\affiliation{Departament de Física de la Mat\`eria Condensada, Universitat de Barcelona, Martí Franqu\`es 1, 08028 Barcelona, Spain}
\affiliation{University of Barcelona Institute of Complex Systems (UBICS), Universitat de Barcelona, 08028 Barcelona, Spain}

\begin{abstract}
We derive and numerically validate an odd Darcy law for chiral active fluids in porous media, in which an antisymmetric permeability tensor induces a flow transverse to the applied pressure gradient. We identify two distinct mechanisms leading to this odd response: odd viscosity in the fluid or the nonlinear coupling of edge currents to inertia. We then show how the transverse transport depends on porosity, boundary friction, and chirality. 

\end{abstract}
\maketitle

\emph{Introduction}---Rome’s aqueducts and lead-pipe networks attest to the sophistication of ancient hydraulic engineering, but a decisive step toward modern water-supply systems came much later in the less renowned French city of Dijon. In the 1850s, while designing the city’s water network, Henri Darcy investigated flow through sand filters and formulated the phenomenological relation now known as Darcy’s law~\cite{darcyFontainesPubliquesVille1856}:
\begin{equation}
\bm U = -\frac{1}{\eta}\bm{K}\cdot\bm{\nabla}\langle\mathtt p\rangle,
\end{equation}
which relates the volume-averaged (superficial) velocity $\bm U$ to the macroscopic pressure gradient $\bm{\nabla}\langle\mathtt p\rangle$. Here, $\eta$ is the fluid viscosity, while the permeability tensor $\bm{K}$ encodes the effects of pore geometry and connectivity on fluid transport. It may be determined empirically from bulk-flow measurements or predicted from pore-scale models using, for example, homogenization~\cite{torquatoRandomHeterogeneousMaterials2002, hornungHomogenizationPorousMedia1997}, closures based on volume or ensemble averaging~\cite{Sahimi2011, whitakerMethodVolumeAveraging1999}, or approximate hydrodynamic cell models~\cite{sanganiSlowFlowPeriodic1982, kuwabara1959forces}. Today, Darcy’s law underpins models of groundwater flow, oil and gas extraction, soil drainage, filtration, dissolution, and geothermal reservoirs~\cite{nield2006convection, golfier2002ability, freeze1979groundwater,dake1978fundamentals}. In biology, where crowded and porous environments are ubiquitous, it describes interstitial fluid transport in articular cartilage, tumors, the lymphatic interstitium, brain parenchyma, and deformable tissues~\cite{MowEtAl1980, PozrikidisFarrow2003, RooseSwartz2012, SchrederEtAl2022, Cheng2016}. 

Many biological media are also active: cells, molecular motors, and other internally driven components continuously generate stresses and flows~\cite{marchetti2013hydrodynamics, bechinger2016active, AlertTrepat2020, NeedlemanDogic2017, ProstEtAl2015, elgeti2015physics}. Extending Darcy-type descriptions to active media is therefore essential for understanding biological transport, especially because active matter responds strongly to obstacles via boundary accumulation, anchoring-dependent sorting, or active surface stresses~\cite{alonso2019transport, modicaPorousMediaMicrostructure2022, reichhardtActiveMatterTransport2014, x8y2-tw4c, shea2024boundary, bhattacharyya2025active}. Recent progress includes studies of active self-propelled particles~\cite{das2025vorticity} or nematics in Hele-Shaw cells~\cite{mackayDarcysLawFriction2020}, corrugated channels~\cite{vaidya2024active} and porous media~\cite{keoghActiveDarcysLaw2024}, the latter revealing a nonmonotonic dependence of permeability on activity.  Much less is known about transport in chiral systems, ranging from circle-swimming particles and suspensions of spinning colloids to living assemblies of rotating cells~\cite{mecke2023simultaneous,chan2024chiral,katuri2024control,chen2025selfpropulsion,tan2022odd,chao2026selective}. At the hydrodynamic level, chirality permits nondissipative antisymmetric transport coefficients~\cite{maire2026kinetic,lier2026chapman,eren2025collisional,hargus2021odd, markovich2025chiral}, most notably odd viscosity~\cite{fruchart2023odd,hosaka2021nonreciprocal,hosaka2023lorentz,lier2023lift,lier2024odd,daddi2026exact,hosaka2024chirotactic}. Boundaries then play a central role~\cite{chan2024chiral,meng2020transport,irani2022dynamics,van2022role,siebers2023exploiting, mecke2025obstacle, mecke2026confinementinducedpumpingchiralactive,uhlig2026chiralityenhancesdiffusiondisordered}: in 2D incompressible fluids, odd viscosity enters the linear response only through stress-dependent boundary conditions~\cite{ganeshan2017odd,lier2024slip}, while torque densities generated by internal spin or transverse chiral forces drive edge currents~\cite{metzger2026equationstateedgeflow,yashunsky2022chiral,kuroda2026designing,Marconi2026hydrodynamics,caporusso2024phase,caprini2025active,wang2026edge,huang2025anomalous, petrini2026curvature, burekovic2026bulk,reyes2026steady}.

Porous media, where fluids repeatedly interact with internal boundaries, provide a natural setting for converting odd stresses into macroscopic transport. Hall-like transport has been observed in active spinner fluids flowing through fixed obstacle arrays~\cite{lou2022odd}. Related transverse responses have also been derived for parity-odd fluids confined in Hele-Shaw geometries~\cite{reynolds2022hele}. These observations raise a broader question. In porous media, how do porosity, odd viscosity, and edge currents affect the transport in a chiral active fluid? We show that Darcy's law remains valid, but with a permeability tensor that acquires an antisymmetric component. We derive this odd permeability quantitatively and identify two distinct origins of the transverse response: odd viscosity and edge currents at finite inertia.

\begin{figure*}
    \centering
    \includegraphics[width=0.999\linewidth]{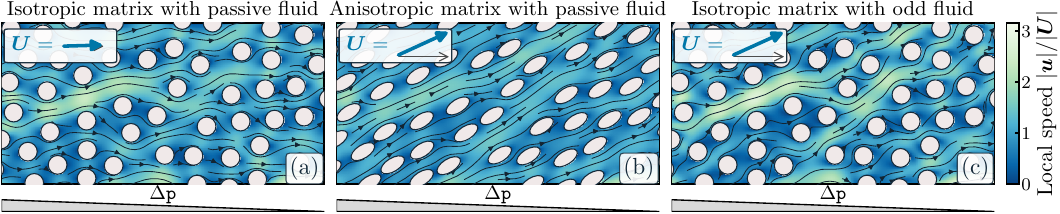}
    \caption{Passive and chiral fluids driven along $\hat{\bm e}_x$ in porous media. Passive flow ($\eta_o=0$) in isotropic (a) and anisotropic (b) media. (c) Chiral flow ($\eta_o/\eta=-1.5$) in an isotropic medium. The insets show the mean flow direction. In all cases $\xi=0$.}
    \label{fig:hero}
\end{figure*}

\emph{Chiral hydrodynamics}---Incompressible 2D chiral fluids are described on mesoscopic scales by~\cite{maire2026kinetic,banerjee2017odd}:
\begin{equation}
    \rho(\partial_t\bm u + \bm u\cdot\bm\nabla\bm u) = \bm \nabla\cdot \bm \Sigma, \qquad \bm\nabla\cdot\bm u = 0,
\end{equation}
with $\rho$ the mass density and $\bm u(\bm r,t)$ the local velocity field driven by the stress:
\begin{equation}
    \Sigma_{ij}\!=\!-\mathtt p\delta_{ij} + \tau\varepsilon_{ij}\!+\eta(\partial_i u_j+\partial_j u_i)+\eta_o(\varepsilon_{jk}\partial_i  u_k+\varepsilon_{ik}\partial_k u_j),
    \label{eq:stress}
\end{equation}
Here, repeated indices are summed, with $\varepsilon_{xy}=-\varepsilon_{yx}=1$ and $\varepsilon_{xx}=\varepsilon_{yy}=0$. The pressure $\mathtt p$ enforces incompressibility, while an imposed pressure difference across opposite boundaries drives a flow. $\eta$ and $\eta_o$ denote the even and odd viscosities, respectively, the latter requiring broken parity and time-reversal symmetries. The homogeneous torque density $\tau(\rho)$ contributes an antisymmetric stress characteristic of chiral systems, arising from effective transverse interparticle forces or the elimination of internal spin~\cite{maire2026kinetic, Marconi2026hydrodynamics}. As the transverse counterpart of pressure, it exerts tangential forces at boundaries that drive edge currents.

We model the porous medium as a collection of fixed obstacles, representing walls or large colloids in spinner suspensions, cells in biological tissues, or micropillars in microfluidic arrays. At their surfaces $\partial \mathcal B$, we impose impermeability and Navier-slip boundary conditions.
\begin{equation}
\begin{aligned}
\bm u(\bm r)\cdot\hat{\bm n}(\bm r) &= 0,\\
\hat{\bm t}(\bm r)\cdot\bm\Sigma(\bm r)\cdot\hat{\bm n}(\bm r) &= \xi\bm u(\bm r)\cdot\hat{\bm t}(\bm r),
\end{aligned}
\qquad\bm r\in\partial\mathcal B.
\label{eq:BC_main}
\end{equation}
where $\hat{\bm n}$ points from the solid into the fluid, $\hat{\bm t}=-\bm\varepsilon\cdot\hat{\bm n}$ is the corresponding tangent, and $\xi$ is the wall-friction coefficient. The limits $\xi\to0$ and $\xi\to\infty$ recover perfect slip and no slip ($\bm u(\bm r)\cdot\hat{\bm t}(\bm r)=0$), respectively.

For constant $\eta_o$ in 2D, incompressibility reduces the odd-viscous force density to
\begin{equation}
\bm\nabla\cdot\bm\Sigma_{\rm odd}=\eta_o\bm\nabla\omega,\qquad \omega=\varepsilon_{ij}\partial_i u_j,
\end{equation}
which can be absorbed into the effective pressure $\mathtt p_{\mathrm{eff}}=\mathtt p-\eta_o\omega$. Odd viscosity therefore affects the flow only through the boundary conditions. It is invisible under no-slip, where the boundary velocity is fixed, but enters under partial slip through the tangential traction. Finite slip also permits the edge currents suppressed by no-slip boundaries, yielding richer and more realistic dynamics.

\emph{Odd viscosity dominated Stokes regime}---We first consider steady Stokes flow, $\bm\nabla\cdot\bm\Sigma=0$, without edge currents ($\tau=0$), such that chirality enters exclusively through the odd viscosity. To model \emph{an extended} porous medium without external boundary effects, we periodically repeat a unit cell containing obstacles. In Fig.~\ref{fig:hero}, we solve the resulting steady-state equations using a Taylor-Hood finite-element discretization (see End Matter), with the flow driven by a horizontal pressure drop and the porous medium represented by obstacles where the boundary conditions in Eq.~\eqref{eq:BC_main} apply. Panel~(a) shows an achiral fluid ($\eta_o=0$) driven through an isotropic medium by a pressure drop $\Delta\mathtt p$ between the boundaries. The resulting mean velocity, $\bm U\simeq U\hat{\bm e}_x$, is aligned with the forcing. In panel~(b), the fluid remains passive, but anisotropy of the medium deflects the mean flow away from the forcing direction. Panel~(c) exhibits a comparable transverse response, but now in an isotropic medium, where it is generated solely by odd viscosity in the fluid ($\eta_o\neq0$). Although their macroscopic signatures may appear similar, geometric anisotropy of the porous medium and odd viscosity of the fluid are fundamentally different mechanisms for producing transverse flow. The anisotropy-induced deflection depends on the orientation of the medium and vanishes when the pressure gradient is aligned with a principal axis. By contrast, the transverse response in the isotropic medium reflects the fluid's chirality, here manifested through odd viscosity. As we show later, it can also arise from other chiral effects, such as torque-driven edge currents coupled to inertia. We first focus on the odd-viscous mechanism.

\begin{figure*}
    \centering
    \includegraphics[width=0.999\linewidth]{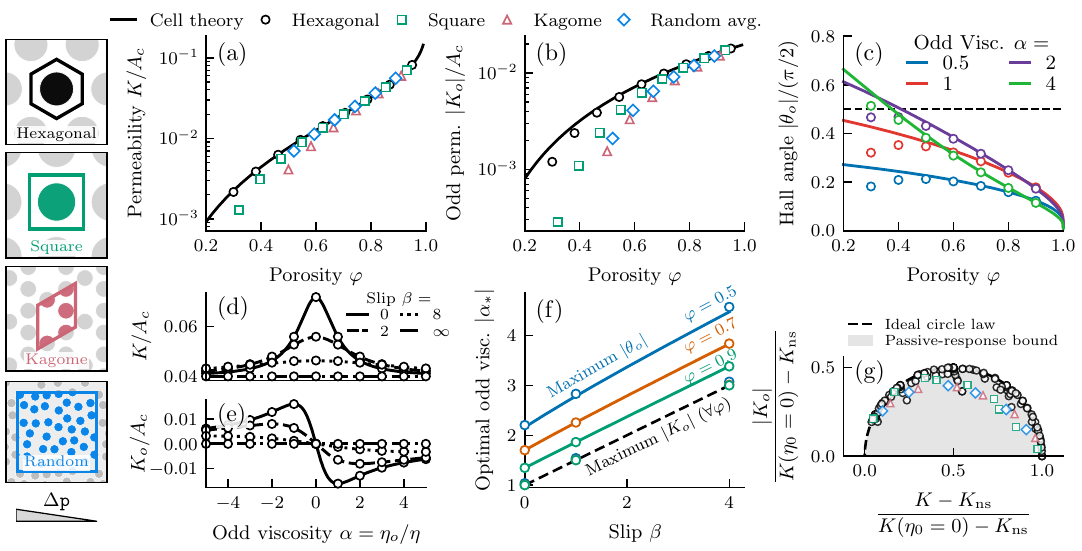}
    \caption{Comparison of theory and simulations for periodic porous media. (a,b) Even and odd permeabilities versus porosity for several lattices and periodic arrangements of randomly positioned disks averaged over 3 realizations, with $|\alpha|=1$ and $\beta=0$. (c) Hall angle $\theta_o=\arctan(K_o/K)$ versus porosity and odd viscosity for the hexagonal lattice at $\beta=0$. (d,e) Even and odd permeabilities versus $\alpha$ and the slip parameter $\beta$ for the hexagonal lattice at $\varphi=0.9$. (f) Values of $\alpha$ maximizing $|\theta_o|$ and $|K_o|$, compared to the hexagonal lattice. (g) Normalized response for all simulated geometries, porosities, and slip coefficients. The gray disk is the passive-response bound Eq.~\eqref{eq:pas}. Opposite signs of $\alpha$ are folded through $|K_o|$. Here $A_c=\pi a^2/(1-\varphi)$.}
    \label{fig:perm}
\end{figure*}

\emph{Odd viscosity induced odd permeability}--- We consider a periodic medium containing $n$ nonoverlapping circular obstacles of radius $a$ in an area $A$, with porosity $\varphi=1-n\pi a^2/A$. The pressure is decomposed as
\begin{equation}
\mathtt p(\bm r)=p(\bm r)+\bm\nabla\langle\mathtt p\rangle\cdot\bm r,
\end{equation}
where $p(\bm r)$ is the periodic part of $\mathtt p$, while $\bm\nabla\langle\mathtt p\rangle\cdot\bm r$ represents its macroscopic affine component (e.g., $\bm\nabla\langle\mathtt p\rangle=-(\Delta\mathtt p/L)\hat{\bm e}_x$ for an inlet-to-outlet pressure drop $\Delta\mathtt p\equiv\mathtt p(0)-\mathtt p(L)$ across a length $L$). This isolates the boundary-induced pressure drop in the affine term, allowing periodic boundary conditions on $p$ while representing the driving as a uniform body force.

To better understand how the transverse response emerges, we adapt a theory developed for a passive porous medium to our case~\cite{kuwabara1959forces, Lasseux_Valdes-Parada_2024}. To this end, we adapt Kuwabara's circular-cell approximation, replacing the complex porous medium by a single circular obstacle of radius $a$ surrounded by fluid up to an effective radius $R=a/\sqrt{1-\varphi}$ that preserves the porosity. At $r=a$, we impose Eq.~\eqref{eq:BC_main} while at an effective cell boundary $r=R$, with $\hat{\bm e}_r=(\cos\theta,\sin\theta)$, we impose:
\begin{equation}
\bm u(R,\theta)\cdot\hat{\bm e}_r=\bm U\cdot\hat{\bm e}_r,\qquad \omega(R,\theta)=\varepsilon_{ij}\partial_i u_j|_{r=R}=0.
\end{equation}
The first condition fixes the cell flux through the area-averaged velocity $\bm U$. The second is the Kuwabara closure~\cite{kuwabara1959forces}, which effectively accounts for neighboring obstacles in a many-body system by imposing zero vorticity at the cell boundary (see Fig.~\ref{fig:kuwabara}). This approximation is most accurate for well-separated obstacles in locally isotropic environments. Deferring the derivation to the End Matter, the resulting area-averaged velocity obeys Darcy's law:
\begin{equation}
\bm U=-\dfrac{1}{\eta}\bm K\cdot\bm\nabla\langle\mathtt p\rangle,
\label{eq:darcy}
\end{equation}
where the permeability tensor reads
\begin{equation}
\bm K=
\begin{pmatrix}
K_{\mathrm{ns}}+\dfrac{c}{c^2+\alpha^2}\Delta K & \dfrac{\alpha}{c^2+\alpha^2}\Delta K\\
-\dfrac{\alpha}{c^2+\alpha^2}\Delta K & K_{\mathrm{ns}}+\dfrac{c}{c^2+\alpha^2}\Delta K
\end{pmatrix},
\label{eq:S_tensor}
\end{equation}
with $\alpha=\eta_o/\eta$, $\beta=\xi a/\eta$, $c=1+\beta/2$, and $\Delta K=K_{\mathrm{ps}}-K_{\mathrm{ns}}$. Here, $K_{\mathrm{ps}}(\varphi)\equiv K_{xx}(\varphi;\alpha=0,\beta=0)$ and $K_{\mathrm{ns}}(\varphi)\equiv K_{xx}(\varphi;\alpha=0,\beta\to\infty)$ are the perfect-slip and no-slip permeabilities of the corresponding passive fluid. Their dependence on porosity $\varphi$ is given explicitly in the End Matter. The transverse response is encoded by the off-diagonal permeability, which deflects the flow perpendicular to the pressure gradient. Despite this deflection, the permeability remains isotropic, since it is built only from rotation-invariant tensors
\begin{equation}
\bm K\!\equiv\!K\bm 1-K_o\bm\varepsilon,\quad K\!=\!K_{\mathrm{ns}}+\frac{c\Delta K}{c^2+\alpha^2},\quad K_o\!=\!-\frac{\alpha\Delta K}{c^2+\alpha^2},
\end{equation}
where $\bm 1$ is the identity tensor and $K$ and $K_o$ denote the conventional and odd permeabilities, respectively. Eq.~\eqref{eq:darcy} can be recast compactly in complex notation as
\begin{equation}
\mathcal U=\dfrac{\mathcal K}{\eta}\mathcal G,\qquad \mathcal K\equiv K+\mathrm iK_o=K_{\mathrm{ns}}+\dfrac{\Delta K}{1+\beta/2+\mathrm i\alpha},
\end{equation}
with $\mathcal U=U_x+\mathrm iU_y$ and $\mathcal G=-\partial_x\langle \mathtt p\rangle-\mathrm i\partial_y\langle \mathtt p\rangle$. The permeability $\mathcal K$ acquires an imaginary part only when $\alpha=\eta_o/\eta\neq0$, thereby producing a flow component transverse to the pressure gradient. Conversely, for $\alpha=0$, $\mathcal K$ is real, and the transverse response vanishes. The same occurs in the no-slip limit $\beta\to\infty$, where the odd-viscous correction is negligible and $\mathcal K\to K_{\mathrm{ns}}$.

Fig.~\ref{fig:perm} tests the theory across several porous geometries. Panels~(a) and~(b) compare the predicted and simulated permeabilities for periodic lattices and a periodic random array. Agreement is excellent at high porosity, where obstacles are well separated and many-body effects are weak, but worsens as porosity decreases and local packing and lattice anisotropy become important. The hexagonal lattice therefore agrees most closely with the rotationally symmetric cell model, whereas the Kagome lattice exhibits larger deviations despite its global isotropy. Capturing these geometry-specific effects would require solving the Stokes problem in the corresponding unit cell~\cite{sanganiSlowFlowPeriodic1982}. Panel~(c) shows the Hall angle $\theta_o=\arctan(K_o/K)$, which quantifies the transverse deflection of the mean flow. While the cell theory predicts a monotonic dependence on porosity, the simulations reveal a maximum at intermediate porosity, showing that geometry and nontrivial correlations can be used to optimize the transverse response. Panels~(d) and~(e) show the dependence on the odd-viscosity parameter $\alpha$ and slip friction $\beta$. At fixed $\beta$, $K$ decreases monotonically with $|\alpha|$. Its dependence on $\beta$ is more unexpected. For $|\alpha|\leq1$, $K$ decreases monotonically, while for $|\alpha|>1$ it first increases, reaches a maximum at $\beta_*=2(|\alpha|-1)$, and then approaches $K_{\mathrm{ns}}$. Thus, increasing wall friction can enhance longitudinal transport when odd viscosity is sufficiently strong. The odd permeability $|K_o|$ is always suppressed by increasing $\beta$ and is maximal at $|\alpha_*|=1+\beta/2$. Panel~(f) confirms this relation and shows that the value of $\alpha$ maximizing the Hall angle is different and depends strongly on porosity. Similar optima arise in other chiral systems~\cite{hosaka2021hydrodynamic,hosaka2023hydrodynamics,abdoli2020nondiffusive}.

Because odd viscosity does no work, energy balance bounds the odd permeability independently of the porous medium geometry (see SM)~\cite{supp}:
\begin{equation}
    K_o^2\leq (K-K_{\mathrm{ns}})(K(\alpha=0)-K).
    \label{eq:pas}
\end{equation}
Two passive measurements thus bound the transverse response: $|K_o|\leq (K(\alpha=0)-K_{\mathrm{ns}})/2$. The inequality follows by neglecting a nonnegative dissipation term associated with flow carrying no net flux. This term vanishes in Kuwabara's rotationally symmetric cell, but is generally finite in more complex geometries, moving the response inside the disk in Fig.~\ref{fig:perm}(g).

External damping, which captures momentum loss to a substrate and is particularly relevant in biological systems, leaves our results largely unchanged, as shown in the Supplemental Material (SM)~\cite{supp}. Three-dimensional flows are qualitatively different. In 3D, odd viscosity requires anisotropy and remains dynamically relevant even under no-slip conditions. As shown in the SM~\cite{supp}, it produces an odd permeability with a surprising sign dependence on boundary conditions.

\begin{figure}
    \centering
    \includegraphics[width=0.99\linewidth]{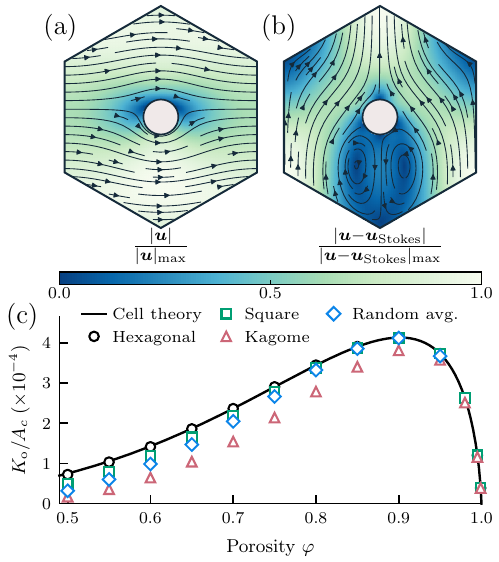}
    \caption{Odd permeability induced by inertia and torque density. (a) Full flow and (b) inertial contribution alone at $\varphi=0.97$. (c) Comparison of measured $K_o$ with theory. Here,  $\alpha=\beta=0$, $\rho|\tau| A_c/\eta^2=1$ and $\rho|\bm G|A_c^{3/2}/\eta^2=1$.}
    \label{fig:inertia}
\end{figure}

\emph{Inertia and torque-driven odd permeability}---We now restore the torque density $\tau$ in Eq.~\eqref{eq:stress}. Since $\tau=\tau(\rho)$ and incompressibility fixes $\rho$, $\tau$ is constant in the fluid bulk. Its only variation is the jump across the fluid-solid interface, so the associated force density $\bm\varepsilon\cdot\bm\nabla\tau$ has support only on $\partial\mathcal B$~\cite{maire2026kinetic2}. Equivalently, the antisymmetric stress acts as a tangential boundary traction, and the boundary condition becomes
\begin{equation}
\hat{\bm t}(\bm r)\cdot\bm\Sigma^{\tau=0}(\bm r)\cdot\hat{\bm n}(\bm r)=\xi\bm u(\bm r)\cdot\hat{\bm t}(\bm r)+\tau(\rho),
\label{eq:driven}
\end{equation}
where $\tau$ has been removed from the bulk stress and retained only as a boundary forcing which drives edge currents~\cite{maire2026kinetic2}. In an isotropic medium, however, their net contributions to a global flow vanish. At the Stokes level, linearity also decouples them from the pressure-driven flow. The Darcy law therefore remains unchanged.

Once inertia is retained, however, the conclusion changes since the \emph{nonlinear} convective term couples the boundary-driven flow to the edge currents. We therefore consider the stationary momentum balance
\begin{equation}
\rho\left(\bm u\cdot\bm \nabla\bm u\right)=\bm \nabla\cdot\bm \Sigma^{\tau=0},
\label{eq:convec}
\end{equation}
subject to impermeability and the driven boundary condition in Eq.~\eqref{eq:driven}. We assume that the pressure gradient and $\tau$ are weak enough such that the stationary solution of Eq.~\eqref{eq:convec} is stable and below the onset of turbulence. We focus on perfect slip, $\beta=0$, and vanishing odd viscosity, with the general case discussed in the SM~\cite{supp}. Figure~\ref{fig:inertia} shows simulations of Eq.~\eqref{eq:convec}. Panel~(a) displays the full flow around an obstacle. Torque-induced circulation weakly breaks the top-bottom symmetry of the pressure-driven flow. Panel~(b) isolates the inertial correction by subtracting the solution without the convective term. The resulting upward flow demonstrates that inertia and edge currents alone generate a transverse response, even without odd viscosity.

To quantify this effect, we solve the cell problem perturbatively to first order in inertia, as detailed in the SM~\cite{supp}. For $\beta=\alpha=0$, the resulting permeabilities are
\begin{equation}
K\simeq\frac{K_{\mathrm{Stokes}}}{1+\Lambda^2},\qquad
K_o\simeq\frac{K_{\mathrm{Stokes}}\Lambda}{1+\Lambda^2},\qquad
\Lambda\propto\frac{\rho\tau a^2}{\eta^2},
\end{equation}
where $K_{\mathrm{Stokes}}$ is the passive-fluid permeability at $\beta=\alpha=0$ found in Eq.~\eqref{eq:S_tensor} in the Stokes regime. The odd permeability is approximately proportional to $\Lambda$, which depends nontrivially on porosity, changes sign with $\tau$, and vanishes with the torque density and its associated edge currents. The theoretical predictions are in good agreement with simulations of Eq.~\eqref{eq:convec} for various lattices as shown in Fig.~\ref{fig:inertia}(c). The nonmonotonic dependence of $K_o$ on porosity results from two competing effects. Increasing porosity reduces viscous resistance but simultaneously decreases the boundary supporting the edge currents. The transverse response is therefore suppressed in both the dense and dilute limits and reaches a maximum at $\varphi \simeq 0.9$ within our cell theory. We also measured in the SM how our first-order approximation degrades as the torque density is increased~\cite{supp}. 

\emph{Conclusion}---We have shown that Darcy's law extends to chiral active fluids through an antisymmetric permeability, and identified two distinct mechanisms underlying this transverse response: odd viscosity, which acts already in the Stokes regime, and torque-driven edge currents, which couple to pressure-driven flow through inertia. Our theory quantitatively predicts these effects and their dependence on porosity and boundary friction.

These findings open several questions about transport in chiral porous media. Relaxing incompressibility may allow the torque density $\tau$ to generate odd permeability already in the Stokes regime. More fundamentally, torque-driven edge currents in porous media remain poorly understood, even in the absence of pressure driving. The effects of weak anisotropy, local porosity variations, and tortuosity also warrant further study, particularly in biological environments. At the opposite extreme, recent studies of turbulence in bulk odd fluids~\cite{de2024pattern,de2026wave,dewit2026oddwaveturbulence} provide a starting point for exploring the largely uncharted turbulent regime of chiral porous media.

\vspace{1.5pt}
\onecolumngrid
\twocolumngrid

\bibliography{bib}

\section*{End Matter}

\emph{Numerical methods}---We solve the steady incompressible momentum equations using a finite-element method. Multi-obstacle periodic systems are meshed with Gmsh~\cite{geuzaine2009gmsh}. The velocity and pressure are discretized using P2-P1 Taylor-Hood elements~\cite{boffi2013mixed}. The macroscopic pressure gradient is represented by a constant body force, while the pressure null mode is removed through a zero-mean constraint~\cite{boffi2013mixed}. Impermeability is imposed as an essential constraint, whereas the slip friction, odd-viscous traction, and imposed torque traction enter through the boundary terms of the weak formulation of the Stokes equation~\cite{boffi2013mixed}. The finite-element integrals in the linear Stokes problem yield a system of equations solved via SciPy~\cite{2020SciPy-NMeth}. When a finite inertia result is wanted, this Stokes solution provides the initial iterate for a Newton method~\cite{kelley2003solving}. The area-averaged velocity $\bm U$ is calculated by integrating the finite-element velocity over the fluid region and dividing by the full system's area.

\emph{Cell theory derivation}---The Stokes equation in the fluid domain is:
\begin{equation}
        -\bm\nabla(p-\eta_o\omega) + \eta\bm\nabla^2\bm u + \bm G = \bm 0.
\end{equation}
where $\bm G = -\bm\nabla \langle \mathtt p\rangle$ and with boundary conditions
\begin{equation}
    \bm u(\bm r)\cdot \hat{\bm n}(\bm r)=0, \quad \hat{\bm t}(\bm r)\cdot\bm\Sigma(\bm r)\cdot\hat{\bm n}(\bm r)=\xi\bm u(\bm r)\cdot\hat{\bm t}(\bm r),
\end{equation}
on $\partial\mathcal B$ where the reduced stress tensor (using $p$ rather than $\mathtt p$), with $\tau=0$ reads:
\begin{equation}
    \Sigma_{ij}=-p\delta_{ij}+\eta(\partial_i u_j+\partial_j u_i)+\eta_o(\varepsilon_{jk}\partial_i  u_k+\varepsilon_{ik}\partial_k u_j).
\end{equation}

\begin{figure}
    \centering
    \includegraphics[width=0.99\linewidth]{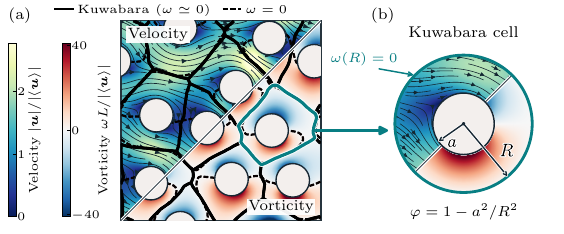}
    \caption{Kuwabara construction with cells around each obstacle corresponding approximately to $\omega=0$. (a) velocity and vorticity fields. The exact points at which $\omega=0$ are also shown, but these are not the ones considered by Kuwabara's construction. (b) Corresponding Kuwabara cell. Here, $|\alpha|=1$ and $\beta=0$.}
    \label{fig:kuwabara}
\end{figure}

In general, such a flow cannot be solved in closed form, even for a circular obstacle in a square box, because the geometry lacks continuous rotational invariance. We therefore study a simpler setup. We suppose that the obstacles are identical, nonoverlapping, and sufficiently evenly spaced, as in Fig.~\ref{fig:kuwabara}. Each obstacle is then assigned a ``cell'' whose outer boundary has approximately zero vorticity (see the construction marked ‘‘Kuwabara'' in Fig.~\ref{fig:kuwabara}(b)). Instead of solving the dynamics of the fluid for the full system, we consider one obstacle of radius $a$ surrounded by an effective circular boundary of radius $R$, representing the cell described above~\cite{kuwabara1959forces}. 

The porosity $\varphi$ of this cell is $\varphi = 1-a^2/R^2$, which can be mapped to the porosity of the original medium. Since a uniform drive excites only the $m=1$ angular harmonic in this rotationally invariant system, the streamfunction can be written as
\begin{equation}
 \psi(r,\theta)=f_x(r)\sin\theta-f_y(r)\cos\theta,
 \label{eq:S_streamfunction}
\end{equation}
with
\begin{equation}
 u_r=\frac{f_x}{r}\cos\theta+\frac{f_y}{r}\sin\theta, \qquad u_\theta=-f_x'\sin\theta+f_y'\cos\theta.
\end{equation}
We package these two real radial amplitudes, and likewise the area-averaged Darcy velocity $\bm U=(U_x, U_y)$ and pressure-gradient drive $\bm G = (G_x, G_y)$, into the complex quantities
\begin{equation}
 f=f_x+\mathrm i f_y,\qquad \mathcal U=U_x+\mathrm i U_y, \qquad \mathcal G = G_x + \mathrm i G_y.
\end{equation}
This is a useful construction to compactly express the boundary conditions and the solution.

Taking the curl of the Stokes equation gives the biharmonic equation $\bm\nabla^4\psi=0$. Both real harmonics obey the same radial equation, so their complex combination has the general solution
\begin{equation}
 f(r)=Ar+\frac{B}{r}+Cr\log\left(\frac rR\right)+Dr^3,
\end{equation}
where $A,B,C,D$ are complex constants.

We now impose the four cell conditions
\begin{align}
 f(a)&=0, \label{eq:S_bc1}\\
 \eta\left[-f''(a)+\frac{f'(a)}{a}\right] +2\mathrm i \eta_o\frac{f'(a)}a+\xi f'(a)&=0, \label{eq:S_bc2}\\
 f(R)&=\mathcal U R, \label{eq:S_bc3}\\
 f''(R)+\frac{f'(R)}R-\frac{f(R)}{R^2}&=0.
 \label{eq:S_bc4}
\end{align}
Eq.~\eqref{eq:S_bc1} is impermeability at the obstacle, and Eq.~\eqref{eq:S_bc2} is the Navier slip condition. At the artificial outer circle $r=R$, Eq.~\eqref{eq:S_bc3} imposes the radial velocity
\begin{equation}
 u_r(R,\theta)=U_x\cos\theta+U_y\sin\theta =\bm U\cdot\hat{\bm e}_r,
\end{equation}
with $\bm U$ conveniently corresponding to the area-averaged velocity. Indeed, incompressibility and impermeability at the obstacle give
\begin{align}
 \frac{1}{\pi R^2}\int_{a<r<R}\bm u \mathrm d\bm r &=\frac{1}{\pi R^2}\int_{r=R}\bm r(\bm u\cdot\hat{\bm n}) \mathrm ds=\bm U.
 \label{eq:cell_mean_velocity}
\end{align}
Eq.~\eqref{eq:S_bc4} is the Kuwabara closure~\cite{kuwabara1959forces}, which imposes vanishing vorticity at $r=R$ to approximate the cancellation of the rotation generated by the central obstacle by that generated in neighboring cells of the original periodic problem, as illustrated in Fig.~\ref{fig:kuwabara}.

Eqs.~\eqref{eq:S_bc1}-\eqref{eq:S_bc4} form the linear system
\begin{equation}
\begin{pmatrix}
a&a^{-1}&a\log(a/R)&a^3\\
L_A&L_B&L_C&L_D\\
R&R^{-1}&0&R^3\\
0&0&2R^{-1}&8R
\end{pmatrix}
\begin{pmatrix}A\\B\\C\\D\end{pmatrix}
=
\begin{pmatrix}0\\0\\\mathcal  U R\\0\end{pmatrix}.
\label{eq:S_linear_system}
\end{equation}
with
\begin{gather}
    L_A=(\eta+2\mathrm i\eta_o)/a+\xi,\quad
    L_B=-(3\eta+2\mathrm i\eta_o)/a^3-\xi/a^2,\nonumber\\
    L_C=\frac{\eta}{a}\log\left(\frac aR\right)+2\mathrm i\frac{\eta_o}{a}[\log\left(\frac aR\right)+1]+\xi[\log\left(\frac aR\right)+1],\nonumber\\
    L_D=(-3\eta+6\mathrm i\eta_o)a+3\xi a^2.
\end{gather}

Inverting the system~\eqref{eq:S_linear_system} yields the solution to the flow. By integrating the traction around the obstacle, the force on it is found. It must be balanced against the cell pressure force $\pi R^2\bm G$, which yields Darcy's law
\begin{equation}
 \mathcal U=\frac{\mathcal K}{\eta}\mathcal G, \qquad \mathcal K = K + \mathrm i K_o, \qquad \mathcal K=\pi R^2\mathcal K^*,
\end{equation}
with
\begin{equation}
 \mathcal K^*(\varphi; \alpha, \beta)=K_{\mathrm{ns}}^*(\varphi)+\frac{K_{\mathrm{ps}}^*(\varphi)-K_{\mathrm{ns}}^*(\varphi)}{1+\beta/2+\mathrm i\alpha},
 \label{eq:S_result}
\end{equation}
where $\alpha={\eta_o}/{\eta}$ and $\beta={\xi a}/{\eta}$ and $K_{\mathrm{ps}}^*$ and $K_{\mathrm{ns}}^*$ are the perfect-slip and no-slip permeabilities in the absence of odd viscosity:
\begin{align}
 K_{\mathrm{ps}}^*&=\frac{-2\log(1-\varphi)-1+(1-\varphi)^2}{16\pi},\\
 K_{\mathrm{ns}}^*&=\frac{-2\log(1-\varphi)-3+4(1-\varphi) -(1-\varphi)^2}{16\pi}.
\end{align}

\clearpage
\onecolumngrid

\begin{center}
\LARGE{\centering\textbf{Supplemental Material}}
\end{center}

\setcounter{page}{1}
\setcounter{equation}{0}
\setcounter{figure}{0}
\renewcommand{\theequation}{S\arabic{equation}}
\renewcommand{\thefigure}{S\arabic{figure}}
\renewcommand{\theHfigure}{supp.\arabic{figure}}
\renewcommand{\theHequation}{supp.\arabic{equation}}
\setcounter{section}{0}
\setcounter{subsection}{0}
\setcounter{subsubsection}{0}
\setcounter{secnumdepth}{3}

\renewcommand{\thesection}{\arabic{section}}
\renewcommand{\thesubsection}{\thesection.\arabic{subsection}}
\renewcommand{\thesubsubsection}{\thesubsection.\arabic{subsubsection}}
\makeatletter

\let\suppoldsection\section
\let\suppoldsubsection\subsection
\let\suppoldsubsubsection\subsubsection

\newcommand{\supplementtoc}{%
  \begingroup
  \setcounter{tocdepth}{3}%
  \def\l@section{\@dottedtocline{1}{0em}{2.8em}}%
  \def\l@subsection{\@dottedtocline{2}{2.8em}{3.6em}}%
  \def\l@subsubsection{\@dottedtocline{3}{6.4em}{4.6em}}%
  \@starttoc{stc}%
  \endgroup
}

\renewcommand{\section}{%
  \@ifstar{\suppoldsection*}{\suppsection}%
}
\newcommand{\suppsection}[1]{%
  \suppoldsection{#1}%
  \addcontentsline{stc}{section}{\protect\numberline{\thesection}#1}%
}

\renewcommand{\subsection}{%
  \@ifstar{\suppoldsubsection*}{\suppsubsection}%
}
\newcommand{\suppsubsection}[1]{%
  \suppoldsubsection{#1}%
  \addcontentsline{stc}{subsection}{\protect\numberline{\thesubsection}#1}%
}

\renewcommand{\subsubsection}{%
  \@ifstar{\suppoldsubsubsection*}{\suppsubsubsection}%
}
\newcommand{\suppsubsubsection}[1]{%
  \suppoldsubsubsection{#1}%
  \addcontentsline{stc}{subsubsection}{\protect\numberline{\thesubsubsection}#1}%
}

\makeatother

\bigskip
\noindent{\large\textbf{Contents}}
\smallskip

\supplementtoc

\bigskip

\section{Energy bound on odd-viscous permeability}
\label{sec:permeability_bound}

We derive Eq.~\eqref{eq:pas} for a pressure gradient $\bm G = -\bm\nabla \langle \mathtt p\rangle =G\hat{\bm e}_x$ and the corresponding flow $\bm u_x$.

For two velocity fields $\bm u$ and $\bm v$, we define their ‘‘mutual dissipation''
\begin{equation}
 \mathcal D(\bm v,\bm u)=2\eta\int_{\Omega_f}\bm E(\bm v):\bm E(\bm u) \mathrm dA +\int_{\partial\mathcal B}\xi v_tu_t \mathrm ds, \qquad E_{ij}(\bm u)=\frac{\partial_i u_j+\partial_j u_i}{2},
 \label{eq:S_bound_dissipation}
\end{equation}
where $\bm E$ is the strain-rate tensor, $u_t=\bm u\cdot\hat{\bm t}$, and $\Omega_f$ is the fluid region of a cell of total area $A$. In particular, $\mathcal D(\bm u,\bm u)\geq0$ is the power dissipated by bulk viscosity and wall friction. 

Let $\bm u_i$ and $\bm u_i^0\equiv \bm u_i(\alpha=0)$ denote the flows driven by $G\hat{\bm e}_i$ with and without odd viscosity, at the same wall friction. Using the no-slip reference $\bm u_i^{\mathrm{ns}}=\bm u_i^0(\beta\to\infty)$, define
\begin{equation}
 \bm w_i^0=\bm u_i^0-\bm u_i^{\mathrm{ns}},\qquad \bm w_i=\bm u_i-\bm u_i^{\mathrm{ns}},\qquad i=x,y.
\end{equation}
The no-slip reference is independent of $\alpha$ in 2D. We study $\bm w_x$, using $\bm w_x^0$ and $\bm w_y^0$ as passive auxiliary fields.

In the steady state, the pressure work is converted into bulk and wall dissipation such that $\mathcal D(\bm u_x,\bm u_x)=AG^2K/\eta$. Multiplying the Stokes equation by a no-slip reference and integrating by parts removes the odd boundary traction and gives $\mathcal D(\bm u_i^{\mathrm{ns}},\bm w_x)=0$. The slip correction therefore dissipates exactly the extra pressure work relative to no slip:
\begin{equation}
 \mathcal D(\bm w_x,\bm w_x)=\frac{AG^2}{\eta}(K-K_{\mathrm{ns}}).
 \label{eq:S_bound_power_identities}
\end{equation}

To identify how much of this power carries mean flow, we use the passive solutions. Multiplying their Stokes equations by an incompressible, periodic, impermeable field $\bm v$ and integrating by parts gives
\begin{equation}
 \mathcal D(\bm u_i^0,\bm v)=G\int_{\Omega_f}v_i \mathrm dA.
 \label{eq:S_bound_passive_work}
\end{equation}
Thus the mutual dissipation with a passive flow measures the flux along its driving direction. By isotropy, $\bm w_x^0$ and $\bm w_y^0$ carry the same extra mean speed $G(K(\alpha=0)-K_{\mathrm{ns}})/\eta$, along $x$ and $y$, respectively. We normalize the extra longitudinal and transverse fluxes by the passive slip-induced flux:
\begin{equation}
 X=\frac{K-K_{\mathrm{ns}}}{K(\alpha=0)-K_{\mathrm{ns}}},\qquad Y=\frac{K_o}{K(\alpha=0)-K_{\mathrm{ns}}}.
\end{equation}
Writing
\begin{equation}
 \bm w_x=X\bm w_x^0+Y\bm w_y^0+\bm w_\perp,
\end{equation}
the remaining motion $\bm w_\perp$ has zero mean flux. Eq.~\eqref{eq:S_bound_passive_work} and the no-slip subtraction give $\mathcal D(\bm w_i^0,\bm w_\perp)=0$, while isotropy gives $\mathcal D(\bm w_x^0,\bm w_y^0)=0$. Eq.~\eqref{eq:S_bound_power_identities} then yields:
\begin{equation}
 K-K_{\mathrm{ns}} =(K(\alpha=0)-K_{\mathrm{ns}})(X^2+Y^2) + \frac{\eta}{AG^2}\mathcal D(\bm w_\perp,\bm w_\perp).
\end{equation}
The first term is the cost of carrying the extra mean flow while the last is dissipation without net transport. Since the latter is nonnegative, by completing the square, we obtain the wanted bond:
\begin{equation}
 \left(X-\frac12\right)^2+Y^2\leq\frac14,\qquad K_o^2\leq(K-K_{\mathrm{ns}})[K(\alpha=0)-K].
\end{equation}
Hence $K_{\mathrm{ns}}\leq K\leq K(\alpha=0)$ and $|K_o|\leq(K(\alpha=0)-K_{\mathrm{ns}})/2$.

Circular-cell theory saturates the bound. Indeed, Eq.~\eqref{eq:S_result}, with $c=1+\beta/2$, gives
\begin{equation}
 X+\mathrm iY=\frac{1}{1+\mathrm i\alpha/c},\qquad \left(X-\frac 1 2\right)^2+Y^2=\frac{1}{4}.
\end{equation}
More complex geometries can generate additional motion with zero mean flux, placing the response inside the disk in Fig.~\ref{fig:perm}(g). The distance from its boundary measures this extra dissipation. 

A fixed substrate damping $\gamma\geq0$ simply adds the nonnegative power $\gamma\int_{\Omega_f}|\bm u|^2 \mathrm dA$ to Eq.~\eqref{eq:S_bound_dissipation} and the argument remains unchanged. By contrast, in the 3D geometry considered below, using the full passive flows instead of the slip corrections gives the weaker bound $K_o^2\leq K_\perp[K(\alpha=0)-K_\perp]$, where $K_\perp=K_{xx}=K_{yy}$. The no-slip subtraction is unavailable because odd viscosity modifies the no-slip flow. In the torque-driven inertial case also considered below, the imposed torque density supplies additional power, so this pressure-work bound no longer applies.

\section{Permeabilities with external damping}

We repeat the calculation done in the End Matter, but now adding damping. The momentum equation is
\begin{equation}
    -\bm\nabla(p-\eta_o\omega)+\eta\bm\nabla^2\bm u -\gamma\bm u+\bm G=\bm 0, \qquad \bm\nabla\cdot\bm u=0,
    \label{eq:S_damped_momentum}
\end{equation}
where $\bm G = -\bm\nabla \langle \mathtt p\rangle$ and the external damping $\gamma$ has units of viscosity per area and defines the screening length $\ell_\gamma=\sqrt{\eta/\gamma}$. We use
\begin{equation}
    \delta=\frac{R}{\ell_\gamma}.
\end{equation}
When $\delta\ll1$, the cell is small compared with the screening length and the undamped result should be recovered (although we must be careful with the Stokes paradox at $\varphi\to 1$). Taking the curl of Eq.~\eqref{eq:S_damped_momentum} removes the pressure and the constant drive and gives
\begin{equation}
    (\bm\nabla^2-\ell_\gamma^{-2})\bm\nabla^2\psi=0,
\end{equation}
with $\psi$ the streamfunction. Using the first-harmonic convention of Eq.~\eqref{eq:S_streamfunction}, the complex radial solution is
\begin{equation}
    f(r)=Ar+\frac Br+C I_1(r/\ell_\gamma)+D K_1(r/\ell_\gamma),
 \label{eq:S_brinkman_basis}
\end{equation}
where $I_1$ and $K_1$ are modified Bessel functions. Define
\begin{equation}
    \mathcal L_1g=g''+\frac{g'}r-\frac{g}{r^2}, \qquad \mathcal T_{\alpha,\beta}g = -g''(a)+\frac{1+\beta+2\mathrm i\alpha}{a}g'(a).
\end{equation}
The four boundary conditions are:
\begin{equation}
    f(a)=0, \qquad \mathcal T_{\alpha,\beta}f=0, \qquad f(R)=\mathcal UR, \qquad \mathcal L_1f(R)=0.
\end{equation}
With
\begin{equation}
    \bm\phi(r)=\bigl(r,r^{-1},I_1(r/\ell_\gamma),K_1(r/\ell_\gamma)\bigr),
\end{equation}
the prescribed-$\mathcal U$ problem is the $4\times4$ system
\begin{equation}
    \begin{pmatrix}
        \bm\phi(a)\\
        \mathcal T_{\alpha,\beta}\bm\phi(a)\\
        \bm\phi(R)\\
        \mathcal L_1\bm\phi(R)
    \end{pmatrix}
    \begin{pmatrix}A\\B\\C\\D\end{pmatrix}
    =\begin{pmatrix}0\\0\\\mathcal UR\\0\end{pmatrix}.
\end{equation}
This determines $A, B, C, D$.

\begin{figure}
    \centering
    \includegraphics[width=0.999\linewidth]{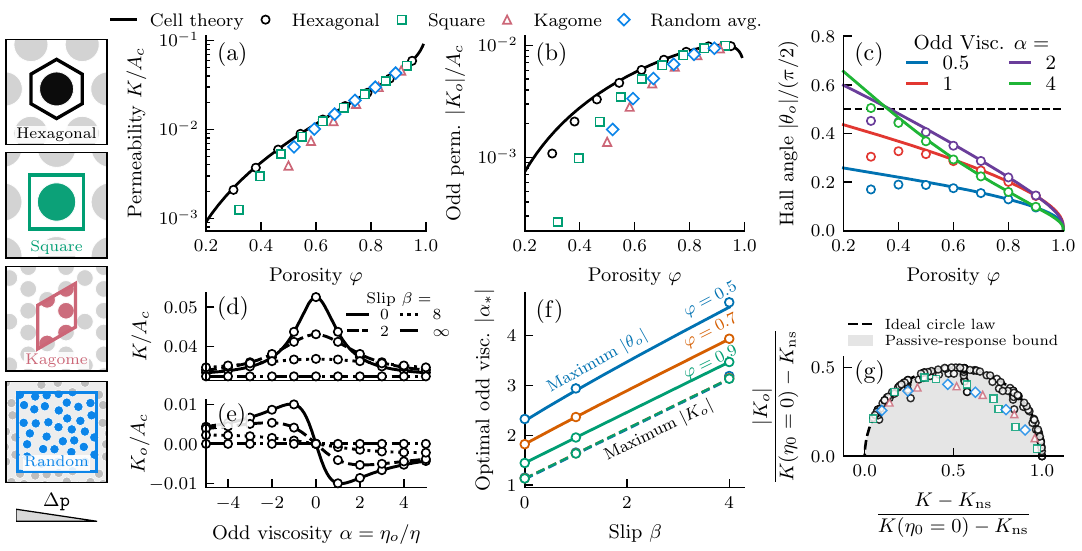}
    \caption{Comparison of theory and simulations for periodic porous media with damping, $\ell_\gamma\simeq0.26$. Panels (a)--(f) show the same quantities as Fig.~\ref{fig:perm}. (g) Normalized response using the passive references of each geometry at the same damping. The dashed semicircle is the circular-cell equality and the gray disk is the passive-response bound.}
    \label{fig:damped}
\end{figure}

Let $\bm F_{\partial\mathcal B}$ be the integral of the complete traction exerted by the fluid on the obstacle, including its pressure, even-viscous, and odd-viscous parts. Force balance yields:
\begin{equation}
    \pi R^2\bm G =\bm F_{\partial\mathcal B} +\gamma\int_{a<r<R}\bm u \mathrm d\bm r = \bm F_{\partial\mathcal B}+\gamma\pi R^2\bm U,
    \label{eq:S_damped_force_balance}
\end{equation}
where the last equality follows from Eq.~\eqref{eq:cell_mean_velocity}. Everything on the right-hand side is known from the prescribed-$\mathcal U$ solution, so Eq.~\eqref{eq:S_damped_force_balance} directly determines $\mathcal G$:
\begin{equation}
    \mathcal U=\frac{\mathcal K_\delta}{\eta}\mathcal G, \qquad \mathcal K_\delta^*\equiv K_\delta^*+\mathrm iK_{o,\delta}^*=K_{{\mathrm{ns}},\delta}^* +\frac{K_{{\mathrm{ps}},\delta}^*-K_{{\mathrm{ns}},\delta}^*}{1+q_\delta(\beta/2+\mathrm i\alpha)}, \qquad \mathcal K_\delta^*=\frac{\mathcal K_\delta}{\pi R^2},
    \label{eq:S_damped_complex_darcy}
\end{equation}
where the endpoint permeabilities and $q_\delta$ have a compact explicit form. We introduce
\begin{equation}
    H_\delta(x)=I_1(\delta x)K_1(\delta)-K_1(\delta x)I_1(\delta)
\end{equation}
and its logarithmic derivative at the obstacle (a prime denotes differentiation with respect to the argument),
\begin{equation}
    h_\delta\equiv\chi\frac{H_\delta'(\chi)}{H_\delta(\chi)}=\chi\delta\frac{I_0(\chi\delta)K_1(\delta)+K_0(\chi\delta)I_1(\delta)}{I_1(\chi\delta)K_1(\delta)-K_1(\chi\delta)I_1(\delta)}-1,\qquad\chi=\frac{a}{R}=\sqrt{1-\varphi}.
\end{equation}
Only the two combinations
\begin{equation}
    A_\delta=2-\varphi+\varphi h_\delta, \qquad B_\delta=\varphi+(2-\varphi)h_\delta
\end{equation}
are needed.  Direct solution of the velocity problem followed by the force balance gives
\begin{equation}
    K_{{\mathrm{ns}},\delta}^*=\frac{A_\delta}{\pi\delta^2 B_\delta},\qquad
    K_{{\mathrm{ps}},\delta}^*=\frac{2A_\delta-\varphi(\chi\delta)^2}{\pi\delta^2\left[2B_\delta-(2-\varphi)(\chi\delta)^2\right]},
    \qquad q_\delta=\frac{2B_\delta}{2B_\delta-(2-\varphi)(\chi\delta)^2}.
\end{equation}
We note that $\chi\delta=a/\ell_\gamma$. Eq.~\eqref{eq:S_damped_complex_darcy} reduces continuously to Eq.~\eqref{eq:S_result} in the undamped limit $\delta\to0$ except at $\varphi\to 1$. For every fixed $\gamma>0$, momentum is screened over the finite length $\ell_\gamma$.  As $\varphi\to 1$ (or $a\to 0$ at fixed $R$), the boundary on which odd viscosity affects the traction disappears, the perfect and no-slip passive permeability converge to the same value $K_{{\mathrm{ps}},\delta}^*, K_{{\mathrm{ns}},\delta}^*\to 1/(\pi\delta^2)$, and hence $K_{o,\delta}^*\to0$.  The remaining uniform flow is simply $\bm U=\bm G/\gamma$.  However, when $\gamma=0$, the momentum is unscreened, and the two-dimensional small-obstacle limit is singular with $K_{\mathrm{ps}}^*-K_{\mathrm{ns}}^*=\varphi^2/(8\pi)$, which remains finite as $\varphi \to 1$, even though the even permeability diverges.  Thus the limits $\gamma\to0$ and $\varphi\to1$ do not commute.

We test our theory in Fig.~\ref{fig:damped} using the same parameters as in the main text, except for a finite damping $\ell_\gamma \simeq 0.26$, and find agreement as good as in the main text. Apart from the behavior of $K_o$ at $\varphi\to 0$, another difference from the undamped case is that now the $\alpha$ at which $|K_o|$ is maximal depends on $\varphi$. Panel~(g) also shows that damping preserves the passive-response bound when the reference permeabilities are evaluated at the same $\gamma$.

\section{Three-dimensional odd permeability}
\subsection{Setup}

We now consider the three-dimensional extension of the cell construction. For simplicity, we do not include a torque density. In 3D, odd viscosity requires a distinguished axis, which we choose to be $\hat{\bm e}_z$. Writing $\epsilon_{ijk}$ for the three-dimensional Levi-Civita symbol, we define $\varepsilon_{ij}\equiv\epsilon_{ij3}$, so that $\bm\varepsilon\cdot\bm v=-\hat{\bm e}_z\times\bm v$, and introduce $\alpha=\eta_o/\eta$. After the same affine-pressure decomposition as in the two-dimensional case, the Stokes equation considered here is
\begin{equation}
    -\bm\nabla p_m+\eta\bm\nabla^2\bm u-\eta_o\hat{\bm e}_z\times\bm\nabla^2\bm u+\bm G=\bm 0,\qquad \bm\nabla\cdot\bm u=0.
    \label{eq:3d_stokes}
\end{equation}
The bulk equation does not uniquely determine the local viscosity tensor in 3D. To evaluate the force on the obstacle, we use the tensor structure of Refs.~\cite{everts2024dissipative,lier2024odd}:
\begin{equation}
    \Sigma_{ij}=-p\delta_{ij}+2\eta E_{ij}+2\eta_o\left(\delta_{ik}\varepsilon_{j\ell}+\varepsilon_{ik}\delta_{j\ell}\right)E_{k\ell},\qquad E_{ij}=\frac{1}{2}(\partial_i u_j+\partial_j u_i).
\end{equation}
The sign of $\eta_o$ is chosen consistently with the two-dimensional stress in Eq.~\eqref{eq:stress}; it is opposite to the convention in those references. The factor of two in the 3D tensor normalization is retained: for an incompressible planar flow, its in-plane odd stress is twice that of Eq.~\eqref{eq:stress} at the same $\eta_o$.
Its divergence gives Eq.~\eqref{eq:3d_stokes}, with $p_m=p-\eta_o\omega_z$ and $\bm\omega=\bm\nabla\times\bm u$. The identity
\begin{equation}
    \hat{\bm e}_z\times\bm\nabla^2\bm u=\partial_z\bm\omega-\bm\nabla\omega_z
\end{equation}
then allows us to rewrite Eq.~\eqref{eq:3d_stokes} as
\begin{equation}
    -\bm\nabla q+\eta\bm\nabla^2\bm u-\eta_o\partial_z\bm\omega+\bm G=\bm 0,\qquad q=p_m-\eta_o\omega_z=p-2\eta_o\omega_z.
    \label{eq:3d_modified_stokes}
\end{equation}
In contrast with the two-dimensional case, odd viscosity modifies the bulk velocity directly, in addition to its contribution to the boundary traction.

\subsection{Cell theory}

We consider a solid spherical obstacle of radius $a$ surrounded by an artificial spherical boundary of radius $R$. We introduce
\begin{equation}
    \chi=\frac{a}{R},\qquad \varphi=1-\chi^3,\qquad \beta=\frac{\xi a}{\eta}.
\end{equation}
At the solid obstacle, we impose impermeability and Navier slip. Defining the tangential projector $\bm P_t=\bm 1-\hat{\bm e}_r\hat{\bm e}_r$ and the tangential velocity $\bm u_t=\bm P_t\cdot\bm u$, the boundary conditions read
\begin{equation}
    \bm u(a)\cdot\hat{\bm e}_r=0,\qquad \bm P_t\cdot\bm\Sigma(a)\cdot\hat{\bm e}_r=\xi\bm u_t(a).
    \label{eq:3d_inner_bc}
\end{equation}
At the artificial outer boundary $r=R$ we impose the radial velocity:
\begin{equation}
    \bm u(R)\cdot\hat{\bm e}_r=\bm U\cdot\hat{\bm e}_r,
\end{equation}
with $\bm U$ again the superficial velocity. We also impose the three-dimensional extension of the Kuwabara closure
\begin{equation}
    \hat{\bm e}_r\times\bm\omega(R)=\bm 0.
    \label{eq:3d_outer_bc}
\end{equation}
The second condition sets the tangential vorticity to zero. For the passive axisymmetric flow, the vorticity is purely azimuthal, and Eq.~\eqref{eq:3d_outer_bc} reduces to the usual Kuwabara condition~\cite{kuwabara1959forces}. The passive construction does not constrain the radial vorticity, which vanishes identically. We therefore do not additionally impose $\hat{\bm e}_r\cdot\bm\omega(R)=0$, because odd viscosity generates a toroidal mode with a non-zero radial vorticity and this extra condition would overconstrain the cell problem.

We first recall the passive solution. Choosing the polar axis along $\bm U$, the streamfunction is written as
\begin{equation}
    \psi_0(r,\theta)=UR^2 F(s)\sin^2\theta,\qquad s={r}/{R}.
\end{equation}
The two inner conditions at $s=\chi$, together with the two outer conditions at $s=1$, become
\begin{equation}
     F(\chi)=0,\qquad \chi F''(\chi)-(\beta+2) F'(\chi)=0,\qquad F(1)=\frac{1}{2},\qquad F''(1)=1.
\end{equation}
Introducing
\begin{equation}
    D_\beta(\chi)=\beta\left(5-9\chi+5\chi^3-\chi^6\right)+3\left(5-6\chi+\chi^6\right),
\end{equation}
we obtain
\begin{equation}
     F(s)=\frac{5(\beta\chi^3+2\beta+6)s^2-15\chi(\beta+2)s+\chi^3\left[5\beta+(6-2\beta)\chi^3\right]s^{-1}-3\chi(\beta+2)s^4}{4D_\beta(\chi)}.
\end{equation}
Integrating the passive traction over the obstacle and balancing it with the pressure force gives the passive spherical Kuwabara permeability:
\begin{equation}
K_{\alpha=0}(\beta)=\frac{2a^2D_\beta(\chi)}{45\chi^3(\beta+2)}=K_{\mathrm{ns}}(\chi)+\frac{K_{\mathrm{ps}}(\chi)-K_{\mathrm{ns}}(\chi)}{1+\beta/2}.
\label{eq:3d_passive_permeability}
\end{equation}
where the 3D perfect-slip and no-slip permeabilities are
\begin{equation}
K_{\mathrm{ps}}(\chi)=\frac{a^2}{15\chi^3}\left(5-6\chi+\chi^6\right),\qquad K_{\mathrm{ns}}(\chi)=\frac{2a^2}{45\chi^3}\left(5-9\chi+5\chi^3-\chi^6\right).
\end{equation}
For $\beta\to\infty$, Eq.~\eqref{eq:3d_passive_permeability} reduces to the usual no-slip spherical Kuwabara permeability~\cite{kuwabara1959forces}.

To include odd viscosity, we expand the velocity and modified pressure as
\begin{equation}
    \bm u=\bm u^{(0)}+\alpha\bm u^{(1)}+\alpha^2\bm u^{(2)}+\mathcal O(\alpha^3),\qquad q=q^{(0)}+\alpha q^{(1)}+\alpha^2q^{(2)}+\mathcal O(\alpha^3).
\end{equation}
The distinguished axis breaks rotational invariance, and the perturbation therefore excites higher spherical-harmonic modes. Although $\mathcal O(\alpha)$ already yields an odd permeability, we retain terms through $\mathcal O(\alpha^2)$, where odd viscosity alone makes the permeability anisotropic.

Eq.~\eqref{eq:3d_modified_stokes} gives the hierarchy
\begin{equation}
    -\bm\nabla q^{(0)}+\eta\bm\nabla^2\bm u^{(0)}+\bm G=\bm 0,\qquad -\bm\nabla q^{(n)}+\eta\bm\nabla^2\bm u^{(n)}=+\eta\partial_z\bm\omega^{(n-1)},\qquad n=1,2.
\end{equation}
The correction fields satisfy homogeneous impermeability and outer conditions. At the obstacle, their tangential boundary conditions are driven by the odd traction generated by the preceding order. We solve this hierarchy using spherical harmonics, separating the case where $\bm U$ is parallel or orthogonal to $\hat{\bm e}_z$. The passive flow is an $\ell=1$ poloidal mode. For a longitudinal flow, the first-order correction is an $\ell=2$ toroidal mode, while for a transverse flow, finite slip also generates an $\ell=1$ poloidal component. The second-order correction contains $\ell=1$ and $\ell=3$ modes. However, only the $\ell=1$ mode contributes to the force applied onto the obstacle by the fluid.

For example, when $\bm U=U\hat{\bm e}_z$, the first-order velocity can be written as
\begin{equation}
    \bm u^{(1)}=-Uh(s)\hat{\bm e}_r\times\bm\nabla_S P_2(\cos\theta),
\end{equation}
where
\begin{equation}
    h(s)=A_{\chi,\beta}\left(s^2+\frac{3}{2}s^{-3}\right)-\frac{H_{\chi,\beta}}{6}s^{-1},
\end{equation}
with
\begin{equation}
    H_{\chi,\beta}=\frac{15\chi(\beta+2)}{2D_\beta(\chi)},\qquad
    A_{\chi,\beta}=\frac{5\chi^3\left[\beta^2+6\beta-8+(12-2\beta)\chi^3\right]}{2D_\beta(\chi)\left[\beta(2\chi^5+3)+12-2\chi^5\right]}.
\end{equation}
Here, $P_2$ is the second Legendre polynomial and $\bm\nabla_S$ is the surface gradient on the unit sphere. The solution satisfies the Navier condition at $s=\chi$ and $[\mathrm d(sh)/\mathrm ds]_{s=1}=0$, but is nonzero in the fluid domain. Odd viscosity therefore generates a transverse bulk flow \emph{even in the no-slip $\beta\to\infty$ limit}.

For compactness, we introduce
\begin{equation}
    C_{\chi,\beta}=\frac{2(1-\chi)(1+\chi+\chi^2)}{3(\beta+2)}
\end{equation}
and
\begin{align}
    X_1=&\frac{15\chi(\beta+6)(1-\chi)^4(1+\chi+\chi^2)(\chi^3+3\chi^2+6\chi+5)}{4D_\beta(\chi)^2},\\
    X_2=&-\frac{75(\beta+6)(1-\chi)^3(1+\chi+\chi^2)^3}{4D_\beta(\chi)^2},\\
    X_3=&\frac{45(\beta+6)(1-\chi)^2(1+\chi+\chi^2)(4\chi^5+4\chi^4+4\chi^3-\chi^2-\chi+5)}{4\chi D_\beta(\chi)^2},
\end{align}
together with
\begin{equation}
    T_{\chi,\beta}=\frac{9}{20}\left(3A_{\chi,\beta}\chi+12A_{\chi,\beta}\chi^{-4}-H_{\chi,\beta}\chi^{-2}\right).
\end{equation}
The three dimensionless resistance coefficients required below are
\begin{align}
    c_o=&\frac{1}{2}-\frac{5(\beta+6)(1-\chi)^2(1+\chi+\chi^2)^2}{2(\beta+2)D_\beta(\chi)},\\
    c_\parallel=&\frac{1}{5}+\frac{3C_{\chi,\beta}}{5}\left[\frac{H_{\chi,\beta}}{\chi}-3A_{\chi,\beta}\left(\chi^2+4\chi^{-3}\right)\right],\\
    c_\perp=&\frac{3}{20}+\frac{\chi D_\beta(\chi)}{45(\beta+2)} \left(X_3-\frac{2X_2}{\chi}-\frac{2X_1}{\chi^2}\right)+C_{\chi,\beta}\left(-\chi T_{\chi,\beta}+\frac{2X_1}{\chi}+\frac{X_2}{2}\right).
\end{align}
Solving the radial equations in the longitudinal and transverse sectors, integrating the traction, and inverting the resulting resistance tensor yields, after some algebra,
\begin{equation}
    \bm K(\alpha,\beta)=K_{\alpha=0}(\beta)
    \begin{pmatrix}
        ~1-\left(c_\perp+c_o^2\right)\alpha^2&-c_o\alpha&0\\[4mm]
        c_o\alpha&1-\left(c_\perp+c_o^2\right)\alpha^2&0\\[4mm]
        0&0&1-c_\parallel\alpha^2
    \end{pmatrix}+\mathcal O(\alpha^3).
    \label{eq:3d_permeability_matrix}
\end{equation}
For $\alpha\neq 0$, an odd permeability emerges in the $xy$ plane. We retain the convention $K_o=K_{yx}=-K_{xy}$ used in the two-dimensional Darcy law. Odd viscosity also induces an anisotropic even response, with $K_{xx}=K_{yy}\neq K_{zz}$, even for an isotropic medium.

\begin{figure}[!ht]
    \centering
    \includegraphics[width=0.98\linewidth]{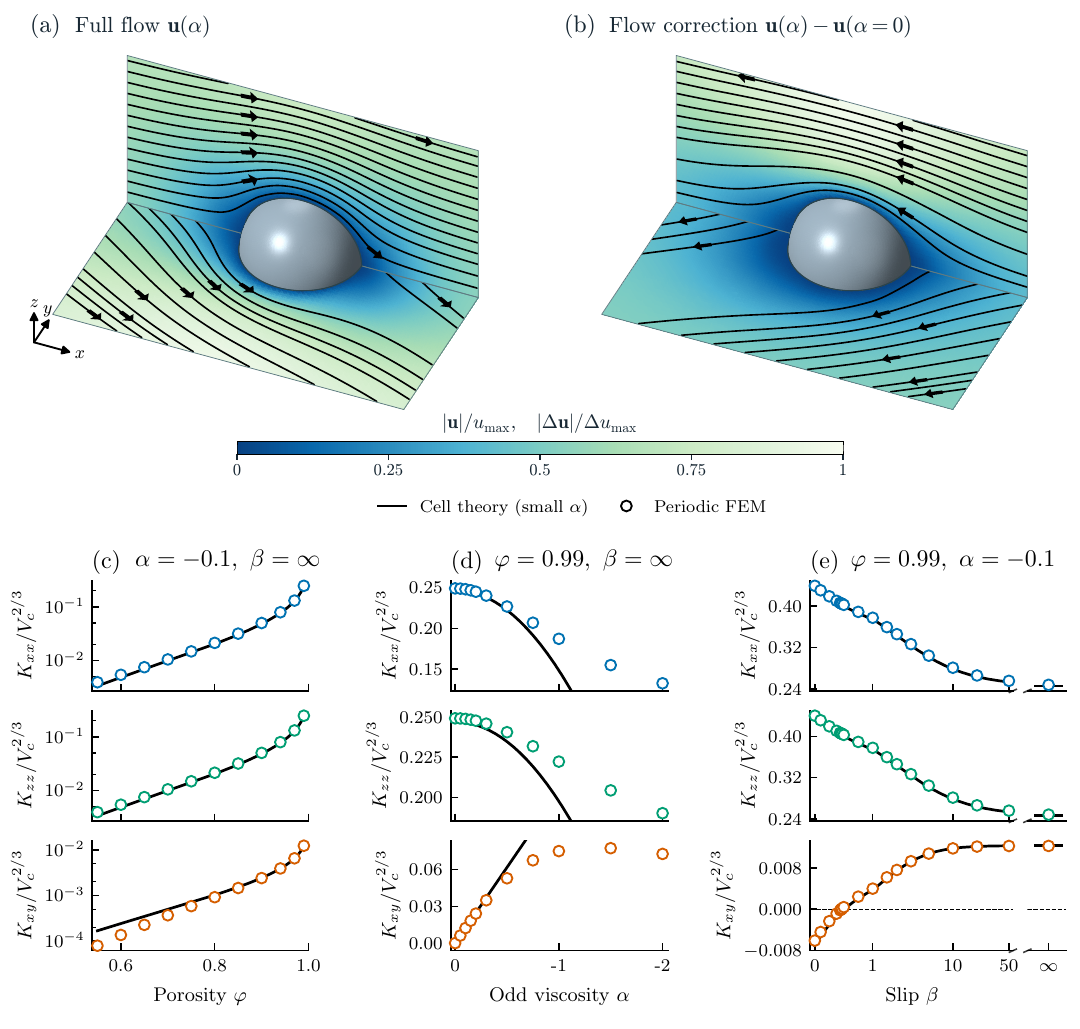}
    \caption{Three-dimensional flow and permeability in a simple-cubic periodic array of spheres, with odd axis $\hat{\bm e}_z$. (a,b) Full flow $\bm u(\alpha)$ and correction $\Delta\bm u=\bm u(\alpha)-\bm u(0)$ at $\varphi=0.99$, $\alpha=-2$, and no slip, driven by $\bm G=G\hat{\bm e}_x$. Colors show $|\bm u|/u_{\max}$ and $|\Delta\bm u|/\Delta u_{\max}$ on the central sections $z=L/2$ and $y=L/2$, respectively, where $u_{\max}=\max_{\bm r}|\bm u(\bm r)|$ and $\Delta u_{\max}=\max_{\bm r}|\Delta\bm u(\bm r)|$ are evaluated over the sampled fluid unit cell. Black curves and arrows follow the velocity projected into each section. The transverse response to this drive is $K_{yx}=-K_{xy}$. (c)--(e) Symbols: periodic simulations; lines: spherical-cell theory, Eq.~\eqref{eq:3d_permeability_matrix}. (c) Porosity scan at $\alpha=-0.1$ and no slip. (d) Odd-viscosity scan at $\varphi=0.99$ and no slip; the perturbative theory is continued over the full range to show its limitations. (e) Friction scan at $\varphi=0.99$ and $\alpha=-0.1$; $\beta=0$ is perfect slip, and the detached $\beta=\infty$ point is an exact no-slip simulation.}
    \label{fig:3d_periodic}
\end{figure}

We can check our results in several limits. For $\alpha=0$, Eq.~\eqref{eq:3d_permeability_matrix} reduces exactly to the passive isotropic result in Eq.~\eqref{eq:3d_passive_permeability}. In the isolated-sphere limit $\varphi\to 1$, it gives the small-$\alpha$ translational mobility of a Navier-slip sphere. Taking subsequently $\beta\to\infty$ where:
\begin{equation}
    c_o\to\frac{1}{2},\qquad c_\perp\to\frac{3}{20},\qquad c_\parallel\to\frac{1}{5},
\end{equation}
recovers, after reversing the sign of their odd-viscosity parameter, the no-slip result of Refs.~\cite{everts2024dissipative,lier2024odd}. The full no-slip limit of the permeability is
\begin{equation}
    \bm K(\alpha, \beta\to\infty)=K_{\alpha=0}(\beta\to\infty)
    \begin{pmatrix}
        1-\dfrac{2}{5}\alpha^2&-\dfrac{\alpha}{2}&0\\[4mm]
        \dfrac{\alpha}{2}&1-\dfrac{2}{5}\alpha^2&0\\[4mm]
        0&0&1-\dfrac{1}{5}\alpha^2
    \end{pmatrix}+\mathcal O(\alpha^3).
    \label{eq:3d_permeability_matrix_no_slip}
\end{equation}
Hence, as expected, in 3D, odd viscosity enters the permeability, even in the no-slip case.

\subsection{Periodic-cell simulations}

We compare the spherical-cell theory with finite-element solutions of Eq.~\eqref{eq:3d_stokes} for a stationary sphere in a cubic periodic cell of side $L$, with $\varphi=1-4\pi a^3/(3L^3)$ and the boundary condition~\eqref{eq:3d_inner_bc}. Both permeabilities are normalized by $V_c^{2/3}=L^2$, matching the total volume $V_c$ of the spherical and periodic cells. Figure~\ref{fig:3d_periodic}(a,b) illustrates the flow at large odd viscosity: a transverse response persists under no slip, while the correction $\bm u(\alpha)-\bm u(0)$ also contains a reduction of the longitudinal flow. For the $+x$ drive shown here, the mean flow is deflected towards $-y$. This Hall response is associated with the antisymmetric $xy$ block, $K_{xy}=-K_{yx}$, while $K_{xz}=K_{zx}=0$. The horizontal section shows the resulting sideways deflection; the $xz$ projection shows local bending around the sphere but no mean deflection along $z$, and omits the transverse $y$ velocity.

At small $|\alpha|$ and high porosity, the theory captures the diagonal and transverse permeabilities [Fig.~\ref{fig:3d_periodic}(c,d)]. Deviations grow at lower porosity, especially for $K_{xy}$, reflecting the approximate spherical outer boundary. The wider $\alpha$ scan also shows the departure from the small-$\alpha$ expansion. Finally, increasing the boundary friction at fixed negative $\alpha$ changes $K_{xy}$ from negative at perfect slip to positive at no slip [Fig.~\ref{fig:3d_periodic}(e)]. Both the periodic simulations and the cell theory predict this sign reversal, so the direction of transverse transport can be reversed by changing slip alone.

\section{Weak inertia and an imposed torque density}

In this section, we derive the permeabilities of a porous medium driven by a pressure gradient combined with a torque density while retaining odd viscosity and finite slip. At the Stokes level, the torque and pressure drivings remain additive because the equations are linear. Once convection, and hence inertia, is included, their cross-advection modifies the permeabilities.

\subsection{Cell problem and torque convention}

We consider the dynamics of an incompressible fluid governed by
\begin{equation}
    \mathcal I\rho(\bm u\cdot\bm\nabla)\bm u=\bm\nabla\cdot\bm\Sigma^{(+)}+\bm G,\qquad \bm\nabla\cdot\bm u=0.
    \label{eq:I_momentum}
\end{equation}
The bookkeeping parameter $\mathcal I$ controls the inertial expansion and may be set to unity at the end of the calculation. The $(+)$ notation is explained below. The reduced stress tensor is
\begin{equation}
    \Sigma^{(+)}_{ij}[\bm u,p]=-p\delta_{ij}+\eta(\partial_i u_j+\partial_j u_i)+\eta_o(\varepsilon_{jk}\partial_iu_k+\varepsilon_{ik}\partial_k u_j). 
\end{equation}
The obstacle boundary conditions are
\begin{equation}
    \bm u(\bm r)\cdot\hat{\bm n}(\bm r)=0,\qquad \hat{\bm t}(\bm r)\cdot\bm\Sigma^{(+)}(\bm r)\cdot\hat{\bm n}(\bm r)=\xi\bm u(\bm r)\cdot\hat{\bm t}(\bm r)+\tau,\qquad \bm r\in\partial\mathcal B.
    \label{eq:bc_torque}
\end{equation}

Before introducing inertia, consider the case without pressure driving, $\bm G=\bm 0$. The torque produces a finite circulating Stokes field, but a system invariant under a nontrivial rotation $\bm R$ cannot acquire a mean velocity from $\tau $ or $\eta_o$:
\begin{equation}
    \langle\bm u_\tau\rangle=\bm R\cdot\langle\bm u_\tau\rangle\quad\Longrightarrow\quad\langle\bm u_\tau\rangle=\bm 0. 
\end{equation}
This applies to the exact square, hexagonal, and Kagome lattices. A finite random realization need not possess this rotational symmetry and may therefore exhibit a realization-dependent torque-induced current, although its isotropic ensemble average must vanish.

\subsection{Stokes solution}

Take the pressure gradient along $x$, $\bm G=G_x\hat{\bm e}_x$. The complete solution in the Stokes regime is the sum of two physical solutions evaluated at the same values of $\eta_o$ and $\xi$:
\begin{enumerate}
    \item $(\bm u_G^{(+)},p_G^{(+)})$ is driven by a unit $x$-directed force and satisfies the homogeneous passive slip condition.
    \item $(\bm u_\tau^{(+)},p_\tau^{(+)})$ has zero body force and unit excess tangential traction.
\end{enumerate}
Both fields satisfy impermeability and periodicity. Linearity of the Stokes problem gives
\begin{equation}
    \bm u^{(0)}=G_x\bm u_G^{(+)}+\tau \bm u_\tau^{(+)}. 
    \label{eq:I_stokes_superposition}
\end{equation}

We then work perturbatively in $\mathcal O(\mathcal I)$ to capture the effects of the inertia. Hence, the nonlinear solutions are expanded as:
\begin{equation}
    \bm u=\bm u^{(0)}+\mathcal I\bm u^{(1)}+\mathcal O(\mathcal I^2),\qquad p=p^{(0)}+\mathcal I p^{(1)}+\mathcal O(\mathcal I^2). 
\end{equation}
Substitution into Eq.~\eqref{eq:I_momentum} gives, at first order,
\begin{equation}
    \bm\nabla\cdot\bm\Sigma^{(+)}[\bm u^{(1)},p^{(1)}]=\rho(\bm u^{(0)}\cdot\bm\nabla)\bm u^{(0)},\qquad \bm\nabla\cdot\bm u^{(1)}=0. 
\end{equation}
The first-order boundary conditions are homogeneous:
\begin{equation}
    \bm u^{(1)}\cdot\hat{\bm n}=0,\qquad \hat{\bm t}\cdot\bm\Sigma^{(+)}[\bm u^{(1)},p^{(1)}]\cdot\hat{\bm n}=\xi\bm u^{(1)}\cdot\hat{\bm t}.
    \label{eq:I_first_order_bc}
\end{equation}

\subsection{Pressure-torque cross-advection}

Using Eq.~\eqref{eq:I_stokes_superposition}, the known first-order forcing is
\begin{equation}
\begin{gathered}
 (\bm u^{(0)}\cdot\bm\nabla)\bm u^{(0)}=G_x^2\bm c_{GG}+G_x\tau \bm c_\times+\tau^2\bm c_{\tau\tau},\\
 \bm c_{GG}=(\bm u_G^{(+)}\cdot\bm\nabla)\bm u_G^{(+)},\qquad \bm c_{\tau\tau}=(\bm u_\tau^{(+)}\cdot\bm\nabla)\bm u_\tau^{(+)},\qquad \bm c_\times=(\bm u_G^{(+)}\cdot\bm\nabla)\bm u_\tau^{(+)}+(\bm u_\tau^{(+)}\cdot\bm\nabla)\bm u_G^{(+)}. 
\end{gathered}
\end{equation}
The pressure-only term is quadratic in $\bm G$, and the torque-only term is independent of $\bm G$, so neither contributes to the linear permeability. The term $\bm c_\times$ is the unique first-order forcing that is linear in both $G_x$ and $\tau$.

To isolate this contribution, define $(\bm u_\times,p_\times)$ by
\begin{equation}
    \bm\nabla\cdot\bm\Sigma^{(+)}[\bm u_\times,p_\times]=\bm c_\times,\qquad \bm\nabla\cdot\bm u_\times=0.
    \label{eq:I_cross_problem}
\end{equation}
This problem satisfies the homogeneous conditions in Eq.~\eqref{eq:I_first_order_bc} and periodicity on the external cell faces. The factor $\rho$ is not included in the definition of $\bm u_\times$ and is restored when the physical first-order field is reconstructed:
\begin{equation}
    \bm u^{(1)}\big|_{G_x\tau }=\rho G_x\tau \bm u_\times. 
    \label{eq:I_cross_mean}
\end{equation}

\subsection{Darcy law and generalized reciprocity}

Using the Lorentz reciprocal theorem adapted to a system with odd viscosity~\cite{hosaka2023lorentz}, we can find the desired result without having to explicitly find $\bm u_\times$.

We define the ‘‘adjoint'' reduced stress by changing the sign of the odd viscous term:
\begin{equation}
    \Sigma^{(-)}_{ij}[\bm v,q]=-q\delta_{ij}+\eta(\partial_i v_j+\partial_j v_i)-\eta_o(\varepsilon_{jk}\partial_iv_k+\varepsilon_{ik}\partial_k v_j).
\end{equation}
For any two incompressible fields $\bm u$ and $\bm v$, the constitutive tensors obey
\begin{equation}
    (\partial_jv_i)\Sigma^{(+)}_{ij}[\bm u,p]=(\partial_ju_i)\Sigma^{(-)}_{ij}[\bm v,q]. 
\end{equation}
The pressure contributions vanish by incompressibility, the even-viscous contributions are symmetric under interchange of the two fields, and the odd-viscous contributions are antisymmetric under that interchange. Integration by parts therefore gives the generalized Lorentz relation
\begin{equation}
    \int_{\Omega_f}\bm v\cdot\bm\nabla\cdot\bm\Sigma^{(+)}[\bm u,p] \mathrm d\bm r-\int_{\Omega_f}\bm u\cdot\bm\nabla\cdot\bm\Sigma^{(-)}[\bm v,q] \mathrm d\bm r=-\int_{\partial\Omega_f}\left[\bm v\cdot\bm\Sigma^{(+)}[\bm u,p]\cdot\hat{\bm n}-\bm u\cdot\bm\Sigma^{(-)}[\bm v,q]\cdot\hat{\bm n}\right] \mathrm ds, 
    \label{eq:I_reciprocal_general}
\end{equation}
where $\Omega_f$ is the fluid domain and, in this boundary integral, $\hat{\bm n}$ points into the fluid domain. On an obstacle, this agrees with the solid-to-fluid normal defined in the main text; the minus sign accounts for this inward-normal convention.

Let $(\bm w_j^{(-)},\pi_j^{(-)})$ be the adjoint passive Stokes field driven by a unit pressure-gradient body force in direction $j$:
\begin{equation}
    \begin{gathered}
    -\bm\nabla\cdot\bm\Sigma^{(-)}[\bm w_j^{(-)},\pi_j^{(-)}]=\hat{\bm e}_j,\qquad \bm\nabla\cdot\bm w_j^{(-)}=0,\\
    \bm w_j^{(-)}\cdot\hat{\bm n}=0,\qquad \hat{\bm t}\cdot\bm\Sigma^{(-)}[\bm w_j^{(-)},\pi_j^{(-)}]\cdot\hat{\bm n}=\xi\bm w_j^{(-)}\cdot\hat{\bm t}. 
    \end{gathered}
    \label{eq:I_auxiliary_j}
\end{equation}
On every obstacle boundary, the integrands in the boundary contributions to Eq.~\eqref{eq:I_reciprocal_general} vanish pointwise. On the external boundary of a periodic cell, opposite faces also cancel pairwise because the fields and reduced stresses are periodic while their outward normals have opposite signs. Using Eqs.~\eqref{eq:I_cross_problem} and \eqref{eq:I_auxiliary_j} therefore gives
\begin{equation}
    \int_{\Omega_f}\bm w_j^{(-)}\cdot\bm c_\times \mathrm d\bm r=-\int_{\Omega_f}u_{\times,j} \mathrm d\bm r. 
\end{equation}
Define
\begin{equation}
    M_j\equiv\dfrac{1}{A}\int_{\Omega_f}u_{\times,j} \mathrm d\bm r=-\frac{1}{A}\int_{\Omega_f}\bm w_j^{(-)}\cdot\bm c_\times \mathrm d\bm r,
    \label{eq:I_reciprocal_M}
\end{equation}
with $A$ the total periodic-cell area, including the fluid and solid regions. Then using Eqs.~\eqref{eq:I_cross_mean} and \eqref{eq:I_reciprocal_M}, we find:
\begin{equation}
    \mathcal U=\frac{\mathcal K_{\mathrm{Stokes}}}{\eta}G_x+\mathcal I\rho G_x\tau \mathcal M+\mathcal O(\mathcal I G_x^2,\mathcal I^2), \qquad \mathcal M\equiv M_x+\mathrm iM_y,\qquad \mathcal K_{\mathrm{Stokes}}\equiv K_{\mathrm{Stokes}}+\mathrm iK_{o,\mathrm{Stokes}},
\end{equation}
where $\mathcal K_{\mathrm{Stokes}}$ is the permeability found in the Stokes regime ($\mathcal I=0$). Comparing the term linear in $G_x$ with $\mathcal U=(\mathcal K/\eta)G_x$ gives
\begin{equation}
    \mathcal K=\mathcal K_{\mathrm{Stokes}}+\mathcal I\rho\eta\tau\mathcal M+\mathcal O(\mathcal I^2),
    \label{eq:I_permeability_correction}
\end{equation}
where pressure-only terms proportional to $G_x^2$ belong to the nonlinear Darcy-Forchheimer response and are excluded from the linear permeability~\cite{Lasseux_Valdes-Parada_2017}. In general, inertial effects are subtle~\cite{10.1063/1.3615514} and here, we only focus on the small contribution from the torque density.

\subsection{Inertial cell theory}

Up to this point, Eq.~\eqref{eq:I_reciprocal_M} is exact for a periodic cell. We now approximate the three zero-inertia fields entering its volume integral by their circular-cell counterparts using Kuwabara construction again. The annular domain contains one obstacle of radius $a$ and an artificial outer boundary of radius $R$, so
\begin{equation}
    \phi_s\equiv1-\varphi=\frac{a^2}{R^2},\qquad A_c=\pi R^2,\qquad s\equiv\frac rR,\qquad \chi\equiv\sqrt{\phi_s}.
\end{equation}
The obstacle is located at $s=\chi$.

\subsubsection{Pressure-driven Stokes flow}

In the End Matter, we solved the pressure-driven Stokes flow in the Kuwabara cell. We recall the steps here in convenient reduced units.
Introduce
\begin{equation}
    z\equiv\beta+2\mathrm i\alpha,\qquad \alpha=\frac{\eta_o}{\eta},\qquad \beta=\frac{\xi a}{\eta},\qquad \mathcal L_s\equiv\log\phi_s.
\end{equation}
For the physical field driven by a unit $x$-directed pressure-gradient body force, write
\begin{equation}
    f_G(r)=\mathcal U_{\mathrm{Stokes}}RF_z(s),\qquad \mathcal U_{\mathrm{Stokes}}\equiv \dfrac{1}{\pi R^2}\int_{a<r<R}\left(u_{G,x}^{(+)}+\mathrm i u_{G,y}^{(+)}\right) \mathrm d\bm r=\frac{\mathcal K_{\mathrm{Stokes}}}{\eta}.
\end{equation}
The four circular-cell conditions are
\begin{equation}
    F_z(\chi)=0,\qquad -F_z''(\chi)+\frac{1+z}{\chi}F_z'(\chi)=0,\qquad F_z(1)=1,\qquad F_z''(1)+F_z'(1)-F_z(1)=0.
\end{equation}
Their solution was already found to be
\begin{equation}
    F_z(s)=\frac{N_z(s)}{D_z},
\end{equation}
with
\begin{equation}
    N_z(s)=2\left[z(1-\phi_s+\mathcal L_s)+2\mathcal L_s\right]s+\phi_s\left[z(\phi_s-2)-2\phi_s\right]\frac1s-4(z+2)s\log s+(z+2)s^3,
\end{equation}
and
\begin{equation}
    D_z=2(1-\phi_s^2+2\mathcal L_s)+z(\phi_s^2-4\phi_s+3+2\mathcal L_s).
\end{equation}
The zero-inertia complex permeability is the previously found
\begin{equation}
    \mathcal K_{\mathrm{Stokes}}^*\equiv\frac{\mathcal K_{\mathrm{Stokes}}}{\pi R^2}=-\frac{D_z}{16\pi(z+2)}=K_{\mathrm{ns}}^*+\frac{K_{\mathrm{ps}}^*-K_{\mathrm{ns}}^*}{1+\beta/2+\mathrm i\alpha}. 
\end{equation}
For the reciprocal integral, it is useful to eliminate $D_z$ between $F_z$ and $\mathcal U$:
\begin{equation}
    \widehat S_z(s)\equiv\frac{\eta f_G(Rs)}{R^3}=-\frac{N_z(s)}{16(z+2)}. 
    \label{eq:I_unit_x_mean}
\end{equation}
The adjoint operator has odd viscosity $-\eta_o$. Rotational invariance and the reality of all other coefficients therefore give the complex radial amplitudes
\begin{equation}
    f_{w_x}(r)=\overline{f_G(r)},\qquad f_{w_y}(r)=\mathrm i\overline{f_G(r)}. 
    \label{eq:I_auxiliary_streamfunction}
\end{equation}

\subsubsection{Torque-driven Stokes flow}

The torque-only Stokes solution is axisymmetric and azimuthal:
\begin{equation}
    \bm u_\tau^{(+)}=h_\beta(r)\hat{\bm e}_\theta,\qquad h_\beta(r)=Ar+\frac Br. 
\end{equation}
Odd viscosity contributes only to the normal traction of this axisymmetric field and does not enter its tangential boundary condition. For the tangent and normal convention used here,
\begin{equation}
    \hat{\bm t}\cdot\bm\Sigma^{(+)}\cdot\hat{\bm n}=-{2\eta B}/{a^2},\qquad \bm u_\tau^{(+)}\cdot\hat{\bm t}=h_\beta(a).
\end{equation}
For the $m=0$ mode, the Kuwabara closure is imposed on the circulation rather than the vorticity. The circulation at the artificial outer boundary is set to zero to reflect the cancellation of the identical rotational flows generated by neighboring obstacles:
\begin{equation}
    \oint_{r=R}\bm u_\tau^{(+)}\cdot\mathrm d\bm\ell=0\Rightarrow A=-{B}/{R^2}. 
\end{equation}
The unit excess-traction condition in Eq.~\eqref{eq:bc_torque} then gives
\begin{equation}
    -\frac{2\eta B}{a^2}=1+\frac{\xi B}{a}(1-\phi_s),\qquad B=-\frac{a^2}{\eta[2+\beta(1-\phi_s)]}.
\end{equation}
The velocity field in this cell geometry is therefore
\begin{equation}
    h_\beta(r)=-\frac{a^2}{\eta[2+\beta(1-\phi_s)]}\left(\frac{1}{r}-\frac{r}{R^2}\right).
    \label{eq:I_periodic_rotlet}
\end{equation}

\subsubsection{Periodic reciprocal integral}

Let $f_G=f_{G,x}+\mathrm if_{G,y}$ and define
\begin{equation}
    \mathcal A_\times(r)=h_\beta\left(\frac{2f_G'}r-\frac{f_G}{r^2}\right),\qquad \mathcal B_\times(r)=\frac{f_Gh_\beta'}r-\frac{h_\beta f_G'}r+\frac{h_\beta f_G}{r^2}.
\end{equation}
Writing $\mathcal A_\times=\mathcal A_x+\mathrm i\mathcal A_y$ and $\mathcal B_\times=\mathcal B_x+\mathrm i\mathcal B_y$, the pressure-torque cross-advection is
\begin{equation}
    c_{\times,r}=\mathcal A_x\sin\theta-\mathcal A_y\cos\theta,\qquad c_{\times,\theta}=\mathcal B_x\cos\theta+\mathcal B_y\sin\theta.
\end{equation}
Substitution of the adjoint amplitudes in Eq.~\eqref{eq:I_auxiliary_streamfunction} into Eq.~\eqref{eq:I_reciprocal_M}, followed by the angular integration, gives the complex response
\begin{equation}
    \mathcal M=-\frac{\mathrm i}{R^2}\int_a^R r\left(\frac{f_G\mathcal A_\times}{r}+f_G'\mathcal B_\times\right) \mathrm dr.
    \label{eq:I_cell_M}
\end{equation}

Define
\begin{equation}
    d_\beta\equiv2+\beta(1-\phi_s),\qquad H(s)\equiv\frac1s-s. 
\end{equation}
Using Eqs.~\eqref{eq:I_unit_x_mean} and \eqref{eq:I_periodic_rotlet}, Eq.~\eqref{eq:I_cell_M} becomes
\begin{equation}
    \mathcal M=\mathrm i\frac{\phi_s R^4}{\eta^3d_\beta}\mathcal J_z, \qquad 
    \mathcal J_z=\int_\chi^1\left[\widehat S_z\widehat S_z'H'-H(\widehat S_z')^2+\frac{3H}{s}\widehat S_z\widehat S_z'-\frac{H}{s^2}\widehat S_z^2\right]  \mathrm ds.
    \label{eq:I_periodic_integral}
\end{equation}
The products in Eq.~\eqref{eq:I_periodic_integral} are ordinary complex products, not modulus squares. Since $N_z$ is a linear combination of $s$, $s^{-1}$, $s\log s$, and $s^3$, the integral is elementary:
\begin{equation}
    \mathcal J_z=-\frac{\mathcal P(\phi_s,z)}{96(z+2)^2},\qquad \mathcal P(\phi_s,z)=P_0(\phi_s)+zP_1(\phi_s)+z^2P_2(\phi_s). 
\end{equation}
The three real functions are
\begin{equation}
    \begin{split}
        P_0(\phi_s)&=6(\log\phi_s)^2-6\log\phi_s-16+12\phi_s-3\phi_s^2-24\phi_s^2\log\phi_s+4\phi_s^3+3\phi_s^4\\
        P_1(\phi_s)&=6(\log\phi_s)^2-30\phi_s\log\phi_s-3\phi_s^4+13\phi_s^3-12\phi_s^2+27\phi_s-25\\
        4P_2(\phi_s)&=6(\log\phi_s)^2+24\phi_s^2\log\phi_s-60\phi_s\log\phi_s+6\log\phi_s+3\phi_s^4-26\phi_s^3+51\phi_s^2-6\phi_s-22
    \end{split}
\end{equation}

Therefore,
\begin{equation}
    \mathcal M
    =-\mathrm i\frac{\phi_s R^4}{96\eta^3d_\beta}
    \frac{\mathcal P(\phi_s,z)}{(z+2)^2}.
\end{equation}
The normalized complex permeability is, to first order in inertia,
\begin{equation}
    \mathcal K^*= \mathcal K_{\mathrm{Stokes}}^*+\delta\mathcal K_{\mathrm I}^*+\mathcal O(\mathcal I^2),
\end{equation}
where
\begin{equation}
    \delta\mathcal K_{\mathrm I}^*\equiv\frac{\mathcal I\rho\eta\tau}{\pi R^2}\mathcal M=-\mathrm i C_{\phi_s,\beta}R_{\phi_s,z},
\end{equation}
with
\begin{equation}
    C_{\phi_s,\beta}\equiv\frac{\phi_s\Theta}{96\pi^2d_\beta},\qquad R_{\phi_s,z}\equiv\frac{\mathcal P(\phi_s,z)}{(z+2)^2},\qquad\Theta\equiv \frac{\mathcal I\rho\tau A_c}{\eta^2}.
\end{equation}
Thus, the controlled first-order even and odd permeabilities are
\begin{equation}
    K^*=\operatorname{Re}\mathcal K_{\mathrm{Stokes}}^*+C_{\phi_s,\beta}\operatorname{Im}R_{\phi_s,z}+\mathcal O(\mathcal I^2),
    \qquad
    K_o^*=\operatorname{Im}\mathcal K_{\mathrm{Stokes}}^*-C_{\phi_s,\beta}\operatorname{Re}R_{\phi_s,z}+\mathcal O(\mathcal I^2).
    \label{eq:I_final_components}
\end{equation}

\subsubsection{Resummed resistance closure}
It is possible to ameliorate the perturbative result of Eq.~\eqref{eq:I_final_components} by considering the resistance, rather than the permeability, and expanding it at first order:
\begin{equation}
    \frac{1}{\mathcal K^*}=\frac{1}{\mathcal K_{\mathrm{Stokes}}^*}-\frac{\delta\mathcal K_{\mathrm I}^*}{(\mathcal K_{\mathrm{Stokes}}^*)^2}+\mathcal O(\mathcal I^2).
\end{equation}
We can then obtain back the permeability by inverting this result at this order:
\begin{equation}
    \mathcal K_{\mathrm{res}}^*\equiv\left[\frac{1}{\mathcal K_{\mathrm{Stokes}}^*}-\frac{\delta\mathcal K_{\mathrm I}^*}{(\mathcal K_{\mathrm{Stokes}}^*)^2}\right]^{-1}=\frac{\mathcal K_{\mathrm{Stokes}}^*}{1-\Gamma_{\phi_s,z}},\qquad
    \Gamma_{\phi_s,z} \equiv\frac{\delta\mathcal K_{\mathrm I}^*}{\mathcal K_{\mathrm{Stokes}}^*}=\mathrm i\frac{\phi_s\Theta}{6\pi d_\beta}\frac{\mathcal P(\phi_s,z)}{(z+2)D_z}.
    \label{eq:I_relative_inertial_correction}
\end{equation}
By construction, the resummed expression in Eq.~\eqref{eq:I_relative_inertial_correction} and the direct expansion in Eq.~\eqref{eq:I_final_components} agree through $\mathcal O(\mathcal I)$. The resummed form is an extrapolation beyond controlled first order, but it remains better behaved at larger inertia and yields better numerical agreement.

Hence, the permeabilities are
\begin{equation}
    \mathcal K_{\mathrm{Stokes}}^* =K_{\mathrm{Stokes}}^*+\mathrm iK_{o,\mathrm{Stokes}}^*,\qquad\Gamma_{\phi_s,z}=\Gamma_R+\mathrm i\Gamma_I,
\end{equation}
the resummed components are
\begin{equation}
    K_{\mathrm{res}}^*=\frac{K_{\mathrm{Stokes}}^*(1-\Gamma_R)-K_{o,\mathrm{Stokes}}^*\Gamma_I}{(1-\Gamma_R)^2+\Gamma_I^2},\qquad
    K_{o,\mathrm{res}}^*=\frac{K_{o,\mathrm{Stokes}}^*(1-\Gamma_R)+K_{\mathrm{Stokes}}^*\Gamma_I}{(1-\Gamma_R)^2+\Gamma_I^2}.
\end{equation}

For $\alpha=\beta=0$, which is the case studied in the Letter,
\begin{equation}
    \mathcal K_{\mathrm{res}}^*=\frac{K_{\mathrm{Stokes}}^*}{1-\mathrm i\Lambda}, \qquad \Lambda=\frac{\phi_s P_0(\phi_s)}{48\pi\left[1-\phi_s^2+2\log(\phi_s)\right]}\Theta.
\end{equation}

\begin{figure}
    \centering
    \includegraphics[width=0.99\linewidth]{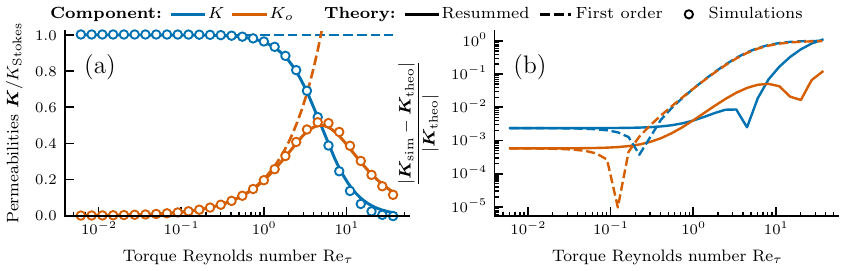}
    \caption{(a) Finite-drive permeability estimates, normalized by the circular-cell $K_{\mathrm{Stokes}}$, and (b) relative error against theory. At fixed $\tau$, the simulations use $G_x=\pm G$ and extract $(K,K_o)=\eta[\bm U(+G)-\bm U(-G)]/(2G)$. The horizontal axis is $\mathrm{Re}_{\tau}=\rho U_{\tau,\mathrm{rms}}\sqrt{A_c}/\eta$, with $U_{\tau,\mathrm{rms}}=[A_c^{-1}\int_{\Omega_f}|\overline{\bm u}_{\tau}|^2\,\mathrm d\bm r]^{1/2}$ and $\overline{\bm u}_{\tau}=[\bm u(+G)+\bm u(-G)]/2$, an estimate of the torque-only flow. The lattice is hexagonal, $\varphi=0.90$, $\alpha=\beta=0$, and $\rho G A_c^{3/2}/\eta^2\simeq0.0403$.}
    \label{fig:inertia2}
\end{figure}

Fig.~\ref{fig:inertia2} shows how the theory deteriorates as the torque increases and how the resumed theory improves upon the naive first-order result. We recall that the simulations are done by enforcing a stationary chiral Navier-Stokes equation. At large Reynolds numbers, the resulting permeability may therefore differ from that obtained in time-dependent simulations, where turbulence may develop.

\subsection{On the use of the reciprocal relation}

The reciprocal-theorem expression for the permeability is derived under periodic boundary conditions, which ensure that the boundary integral vanishes. Our cell geometry is not periodic, so this contribution cannot be discarded a priori. Nevertheless, direct evaluation of the boundary terms shows that it vanishes in this case.

\end{document}